\documentclass[aps,groupedaddress]{revtex4-1}
                       \usepackage{fullpage}
                       \usepackage{amssymb}
                        \usepackage{amsmath}
                        \usepackage{graphicx}
                        \usepackage{mathpazo}
                       \usepackage{dsfont}
                        \usepackage{float}

\begin{document}

\title{Nonlinear waves on circle networks with excitable nodes}
\author{
Shou-Wen Wang
}
\email{wang-sw09@mails.tsinghua.edu.cn}
\affiliation{Department of Engineering Physics, Tsinghua University, Beijing 100084, China}
\author{Yueheng Lan}

\email{lanyh@mail.tsinghua.edu.cn}
\affiliation{Department of Physics, Tsinghua University, Beijing 100084, China}

\date{\today}
                                \begin{abstract}
Nonlinear wave formation and propagation on a complex network with
excitable node dynamics is of fundamental interest in diverse fields
in science and engineering. Here, we propose a new model of 
the Kuramoto type to study nonlinear wave generation and propagation 
on circular subgraphs of a complex network. On
circle networks, in the continuum limit, this model is equivalent to
the over-damped Frenkel-Kontorova model. The new model is shown to
keep the essential features of those well-known models such as the
diffusively coupled B{\" a}r-Eiswirth model but with much simplified
expression such that analytic analysis becomes possible. We classify
traveling wave solutions on circle networks and show the
universality of its features with perturbation analysis and  numerical
computation.
\end{abstract}

\maketitle
\tableofcontents

\section{Introduction}

Synchronization is a collective and emergent behavior of coupled
agents which display spontaneous locking to a common oscillation
frequency. The investigation of this ubiquitous and important
phenomenon has been intense and fruitful in recent years. Much
progress has been made in diverse fields, including examples from
networks of pacemaker cells in the
heart~\cite{peskin1975mathematical,michaels1987mechanisms},
metabolic synchrony in yeast cell
suspensions~\cite{ghosh1971metabolic,aldridge1976cell},
congregations of synchronously flashing
fireflies~\cite{buck1988synchronous,buck1976synchronous}, arrays of
lasers~\cite{jiang1993numerical,kourtchatov1995theory} or microwave
oscillators~\cite{york1991quasi} and wired superconducting Josephson
junctions~\cite{wiesenfeld1996synchronization}. It is probably one
of the best examples of spontaneous emergence of rhythms in
non-equilibrium systems. Different theoretical models have been
proposed to study its onset and stability, among which the Kuramoto
model is the most widely used for its simplicity in formulation and
elegance in analysis. The majority of these models study coupled
oscillators and check how the oscillation of individual members is
shaped by different coupling strengths and topologies. Other
dynamical aspects, such as the impact of noise, the finite-size
effect and the co-evolution of structure and dynamics, have also
been explored~\cite{acebron2005kuramoto,hong2007finite}. Different
types of local dynamics are also explored to various extent with
many interesting observations made~\cite{arenas2008synchronization}
while much more remains to be
probed~\cite{arenas2008synchronization}.

\begin{figure}[ht]
\centering
\includegraphics[width=8cm]{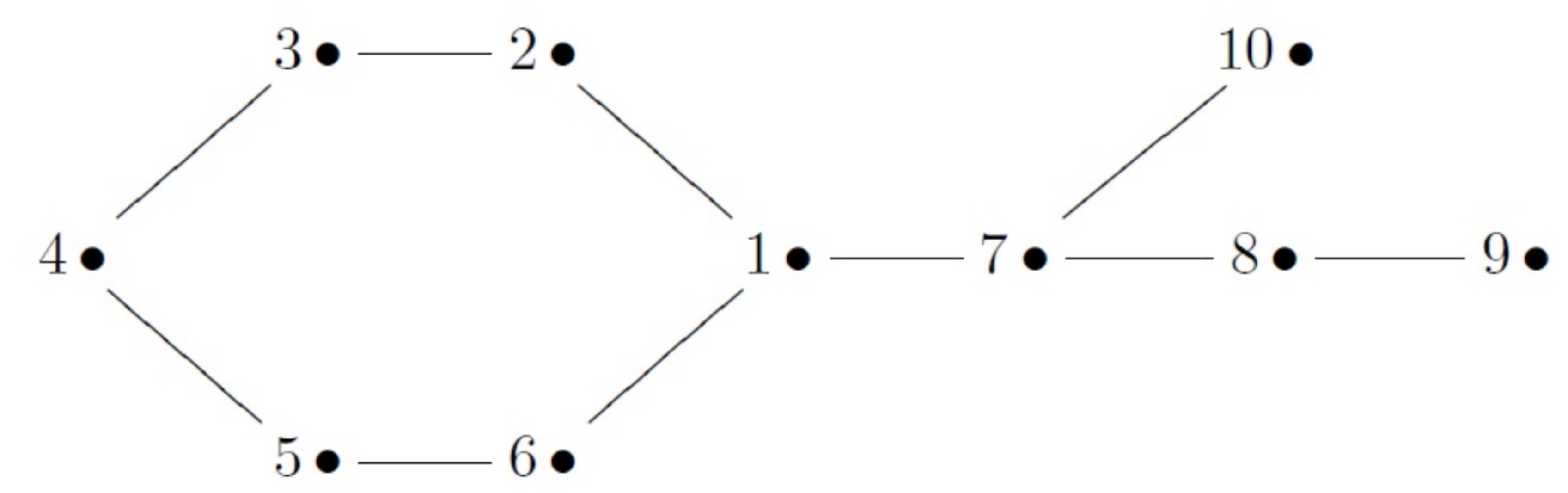}
\caption{A simple example of a network with a circle sub-network.}
\label{fig:circle-sub}
\end{figure}

The spatiotemporal pattern formation in excitable media has long
been a hot topic for researchers in both applied and theoretical
arena. If excitable dynamics is mounted on each node of a network, a
discrete analogue of the excitable media is created, the collective
dynamics of which critically depends on the network structure. Hu
{\em et. al} recently investigated the diffusively coupled B{\"
a}r-Eiswirth model and found that different nonlinear waves may
emerge spontaneously with properly designed coupling
strategy~\cite{qian2010structure}. These nonlinear waves possess an
interesting feature: the evolution on the neighboring nodes is
equally separated in time but not in space. Based on the
comprehensive numerical observation, they proposed an phase-advanced
model which explains how all the nodes of the network are driven by
a self-sustained central circle sub-network of oscillating nodes.
Hence, dynamically a complex network can be viewed in a much simpler
way: the central driving circle sub-network and the attached trees,
which is determined by system dynamics. In
Fig.~\ref{fig:circle-sub}, the nodes 1-6 make the circle subnetwork
and 7-10 are attached as a tree branch. Once a driving circle
network is selected dynamically, the oscillation on it will
determine the behavior of the rest of the network. Therefore, it is
crucial to identify possible spatiotemporal patterns on a circle
network of excitable nodes with diffusive coupling.

In fact, the propagation of nonlinear waves along a circular track
of excitable media was observed experimentally in heart muscles long
ago and has been well explained based on empirical physiology
models~\cite{frame1988oscillations,instability1993}. Similar waves
were also observed in other low dimensional systems, such as the
charge density waves in the quasi 1-d
metals~\cite{gruner1989charge}. In all these studies, a plethora of
wave patterns were explored and their existence and stability were
investigated with different analytical or numerical tools. However,
in their mathematical description, the sophisticated set of coupled
nonlinear differential equations often prevent an analytic approach
and thus hinder our full understanding of the wave dynamics even in
simple cases. Due to the universality of nonlinear wave propagation
on networks, we feel that it is possible and necessary to find a
model which keeps the essential features of those well-known models
such as the diffusively coupled B{\" a}r-Eiswirth model but with
much simplified description such that analytic analysis becomes
possible.

Inspired by the success of the Kuramoto model in synchronization, we
here propose a new model of similar type to target the problem of
spatiotemporal pattern formation in a circle network with excitable
node dynamics. We put a one-dimensional phase oscillator at each
node, which is similar to the Kuramoto model but with nonuniform
local frequency and diffusive coupling. The model is considerably
simpler than the B{\" a}r-Eiswirth model used by Hu {\em et.
al}~\cite{qian2010structure} but captures the essential dynamics.
Specifically, detailed studies on spatiotemporal patterns in circle
network have been carried out. Different spatiotemporal patterns in
circle networks are observed, which can be compared and classified
analytically according to the solution in the limit of uniform local
frequency. Among all the nonlinear waves, the regular nonlinear
wave, characteristic of its equal time-separation, is of great
interest and significance. Further perturbation analysis indicates
that the equal-time-separation solution is ubiquitous in the coupled
excitable dynamics and it is stable. The restitution and dispersion
curves of this wave bear remarkable similarity to those of a ring of
excitable media, which implies universality of our model and its
solutions. Other types of solutions, stable or unstable, are also
found, including one special solution that seems to have no analogue
in the continuum limit.

The equal-time-separation solution is related conceptually to the
lag synchronization of two coupled nonidentical chaotic oscillators,
in which physical observables of the two become synchronized but
with a time lag~\cite{rosenblum1997phase}. However, in our model,
identical excitable phase oscillators are coupled diffusively and
thus both the formed patterns and the underlying interaction between
agents are different. In celestial mechanics, a remotely related
example is the choreographic solution for the n-body problem, in
which the moving bodies are also separated by a constant time
interval~\cite{chenciner2000remarkable}. Henceforth,
this equal-time-separation solution seems to be universal and much
work is needed to reveal its manifestation and implication in
different contexts.

This new model is an over-damped version of the well-known
Frenkel-Kontorova
model~\cite{braun2004frenkel,aubry1983discrete,bak1982commensurate}
which  has become one of the fundamental and universal tools of
low-dimensional nonlinear physics.  The classical Frenkel-Kontorova
model describes a chain of classical particles evolving on the real
line, coupled with their neighbors and subjected to a periodic
potential. In the continuum limit, \emph{i.e.}, the distance $a_0$ between
neighboring nodes satisfies $a_0\ll 1$, the Frenkel-Kontorova model is reduced to
the Sine-Gordon equation, which is a completely integrable nonlinear
partial differential equation.  The simplicity of the
Frenkel-Kontorova model, as well as its surprising richness and
capability to describe a range of important nonlinear phenomena has
attracted a great deal of attention from physicists working in
solid-state physics and nonlinear science, which provides a unique
framework to combine many physical concepts and to make the analysis
in a unified and consistent way. It is hopeful that our simplified
version of the Frenkel-Kontorova model can also give interesting
insights into the pattern formation in the context of complex
networks.

The paper is organized as follows. In section \ref{sect:model}, we
motivate the introduction of our model and a detailed discussion of
its solution is made in section \ref{sect:patt}. In particular, the
stability condition and periods of regular solutions are analyzed
with a perturbation approach. The regular nonlinear wave turns out
to be a generic feature of circle networks with excitable node
dynamics, which is discussed in detail in section \ref{sect:univ}.
The relation of the circulating pulse in the excitable media and the
regular nonlinear wave is investigated in section \ref{sect:corr}
with the dispersion and restitution curves being plotted. We
summarize our results in section \ref{sect:sum}.

\section{Our model}
\label{sect:model}

\begin{figure}[ht]
\centering
\includegraphics[width=10cm]{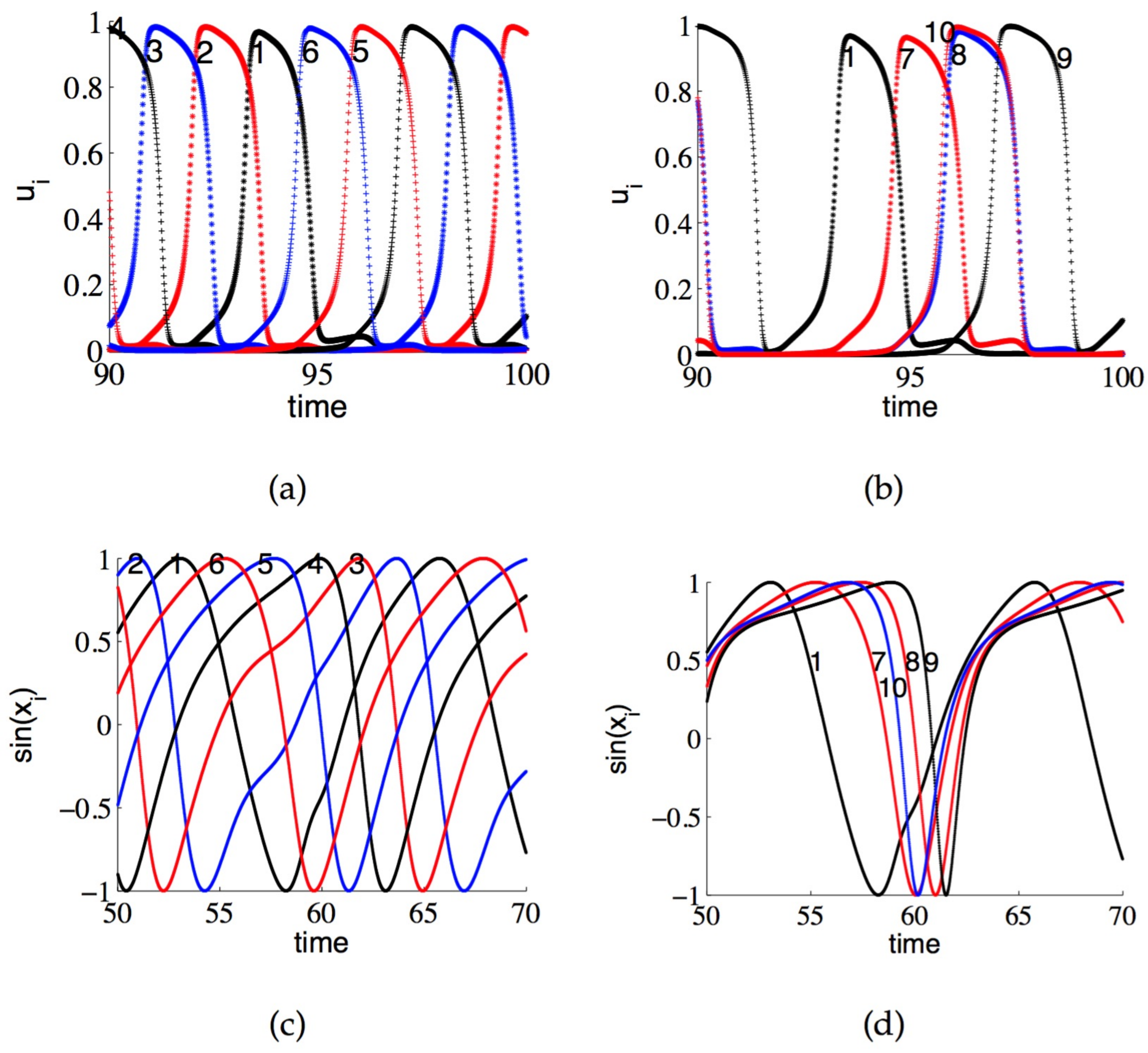}
\caption{Nonlinear waves on the network displayed in
Fig.~\ref{fig:circle-sub}. (a), (b) Dynamics ({\em i.e.}, $u_i(t)$,
plotted as \lq{}*\rq{} or \lq{}+\rq{}) of different nodes for
Eq.~(\ref{eq:F-N-model}) with
$\epsilon=0.04,\,b=0.07,\,a=0.84,\,d=0.2$: for nodes (a) in the
circle sub-network and (b) in the attached tree branch. (c), (d)
Corresponding dynamics for Eq.~(\ref{eq:model1}) with
$w=0.8,\,\epsilon=1,\,d=1$, for nodes (c) in the circle sub-network
and (d) in the branch. Note that the vertical coordinate for (c) and
(d) is $\sin(x_i)$ so that the similarity between (a) (b) and (c)
(d) is more apparent.} \label{fig:two-model-comp}
\end{figure}

Much effort has been devoted to the study of nonlinear wave
propagation or self-sustained oscillation on different types of
networks. The B{\" a}r-Eiswirth model~\cite{BM93} is recently used
by Hu {\em et. al},
\begin{subequations}\label{eq:F-N-model}
\begin{equation}
\frac{du_i}{dt}=-\frac{1}{\epsilon}u_i(u_i-1)( u_i-\frac{v_i+b}{a} ) + \Delta^2u_i\,,
\end{equation}
\begin{equation}
\frac{dv_i}{dt}=f(u_i)-v_i\,,
\end{equation}
\end{subequations}
where
\[
f(u)=
\begin{cases}
0, & \quad u<\frac{1}{3} \,\\
1-6.75u(u-1)^2,&\quad \frac{1}{3}\le u\le 1 \,\\
1,& \quad u>1\,,
\end{cases}
\]
and $\Delta^2$ denotes the discrete Laplacian which can be
defined on a bidirectional graph $D=<V,E>$ with the vertex set
$V=\{v_1,v_2,...,v_n\}$ and the edge set $E=\{e_1,e_2,...,e_m\}$.
With $a_{i,j}$ representing the number of edges from vertex $v_i$ to
vertex $v_j$, the discrete Laplacian is
\begin{equation}
\Delta^2 u_i= \sum_{j=1}^n a_{j,i}(u_j - u_i).
\label{eq:laplacian}
\end{equation}
On a network of similar type as in Fig.~\ref{fig:circle-sub}, with
Eq.~(\ref{eq:F-N-model}), the dominant phase-advanced driving mechanism
(DPAD)~\cite{qian2010structure} governs the dynamics. In
Fig.~\ref{fig:two-model-comp}(a) and \ref{fig:two-model-comp}(b), we
show the dynamical behavior of all nodes if working with the network
shown in Fig.~\ref{fig:circle-sub}. A circulating nonlinear wave is
found on the circle sub-network with nearly uniform velocity, which
indicates the equal time-separation between adjacent nodes. The
small discrepancy is caused by the side branch attached. The
dynamics of the nodes in the side branch is subordinate to that of
node 1, as illustrated in Fig.~\ref{fig:two-model-comp}(b). The DPAD
mechanism, {\em i.e.}, a cascading driving ladder relaying sustained
oscillations, is vividly shown in Fig.~\ref{fig:two-model-comp}(b).
For example, node 7 is always pulled by node 1 away from the stable
fixed point of the local dynamics. The circle sub-network plays an
essential role in the current network which motivates an in-depth
investigation on the possible dynamical behavior of a circle network
with various local dynamics to find out universality of the wave
propagation. Below, we will take one of the simplest case, {\em
i.e.}, a particular one-dimensional equation as the local dynamics.

A general equation of motion with local 1-d dynamics and nearest
neighbor interaction along a circle network could be written as:
\begin{equation}
\dot{x_i}=f(x_i)+R(x_{i-1},x_i,x_{i+1}),  \,\,\,\,i\in \{1,2,3,....,n\}
\,\,,\;x_i\in \mathbb{R} \label{eq:master}
\end{equation}
where $f(x_i)$ denotes the local dynamics and
$R(x_{i-1},x_i,x_{i+1})$ is the coupling term between neighboring
nodes. The total number of nodes is assumed to be $n$ and due to
circle topology, $x_{i\pm n}=x_i$. The dynamics is invariant under
the rotation $x_i \rightarrow x_{i+k}, \forall k \in \mathbb{N} $,
the reflection $x_i \rightarrow x_{n+1-i}$ and their group
composite. The periodicity also implies a simple composition rule.
Suppose the variables $\{x_i(t)\}_{i=1,2,...,p} $ and
$\{y_i(t)\}_{i=1,2,...,mp}$ describe dynamics governed by the same
equation of motion (\ref{eq:master}) on two circle networks with $p$
and $mp$ nodes respectively. The identification $y_i=x_j\,,\mbox{
for }i=j(\mathrm{mod}\, p)\,,i=1,2,...,mp$ will generate a solution
for the $y$-system for any $x$-solution, as illustrated in
Fig.~\ref{fig:regularsolutions}. Dynamics in
Fig.~\ref{fig:regularsolutions}(d) is inherently a spatial
juxtaposition of that in Fig.~\ref{fig:regularsolutions}(g).

For simplicity, without loss of generality, we will mainly use the
diffusion model below
\begin{equation}
\dot{x}_i=\omega-\epsilon
\sin(x_i)+d\sin(\Delta^2 x_i),\,\,\,\,i\in \{1,2,3,....n\} ,\;x_i\in\mathbb{R}
\label{eq:model1}
\end{equation}
where $\Delta^2$ is the Laplacian operator, as defined in
Eq.~(\ref{eq:laplacian}). For  circle networks, $\Delta^2
x_i=x_{i+1}+x_{i-1}-2x_i$.  The local dynamics is identical to that
of the Kuramoto model if $\epsilon=0$. Besides, the case $\epsilon>\omega$ corresponds to the so-called theta-neuron model, also known as Ermentrout-Kopell model~\cite{ermentrout1986parabolic}  and the case $\epsilon<\omega$ is widely investigated in the field of Josephson junctions~\cite{wiesenfeld1998frequency}. When $\epsilon$ is greater
than but close to $\omega$, a stable and an unstable equilibria
exist on the phase circle of the local dynamics, and the system
becomes excitable: all the orbits go to the unique stable
equilibrium unless a perturbation brings the state over the unstable
one which induces a large excursion. Compared to the usual
2-dimensional excitable dynamics equation Eq.~(\ref{eq:F-N-model}),
the current one is much simpler so that a relatively thorough
discussion of its solution becomes possible. In the
continuum limit, the wavelength $\lambda \gg 1$ and $\Delta
x_i=x_{i+1}+x_{i-1}-2x_i\ll 1$, so that $\sin (\Delta^2 x_i) \rightarrow
\Delta^2 x_i$. The wavelength $\lambda$ is measured in terms of the
number of nodes that are spanned with a period of the wave
oscillation. The term $\sin(\Delta^2 x_i)$  is invariant under a
phase shift of $x_i\to x_i+2k_i\pi, k_i\in \mathbb{Z}$, which is
convenient to analyze in the current context.

With Eq.~(\ref{eq:model1}), the dynamical behavior of all nodes in
Fig.~\ref{fig:circle-sub} is displayed in
Fig.~\ref{fig:two-model-comp}(c) and (d). The dynamical behavior in
Fig.~\ref{fig:two-model-comp}(c) looks similar to that in
Fig.~\ref{fig:two-model-comp}(a). Hence, the simple one-dimensional
Eq.~(\ref{eq:model1}) seems to have captured the essential features
of the B{\" a}r-Eiswirth model~(\ref{eq:F-N-model}), though the
precise wave profiles are not identical. Here, we emphasize that
although the interaction between the circle sub-network and the
branch network is mutual, the DPAD mechanism governs the
uni-directional propagation of action, as clearly seen in
Fig.~\ref{fig:two-model-comp}(d). The back reaction of the branch on
the circle network is small, so the equal-time-separation profile is
well preserved. For a general nonlinear wave propagation, this may
not be the case. The observed DPAD structure is intimately related
to the excitable node dynamics. Our model Eq.~(\ref{eq:model1}) well
captures this particular feature of the dynamics.


Our new model is closely related to the Frenkel-Kontorova
model which describes harmonically coupled particle chain moving in
a periodic potential. If $\phi$ denotes the position of the particle
$i\in \mathbb{Z}$, one of the simplest Frenkel-Kontorova models
could be written as
\[m\frac{d^2\phi}{dt^2}+\gamma \frac{d\phi_i}{dt}=L+\sin(\phi_i)+\Delta^2 \phi_i,\]
where $m$ is the particle mass, $\gamma$ a friction coefficient and
$L$ is a constant driving force. The term $\sin (\phi_i)$ is the
force exerted by a periodic potential and the interaction $\Delta^2
\phi_i=\phi_{i+1}+\phi_{i-1}-2\phi_i$ is diffusive.

 In the over-damped limit $m\ll \gamma=1$, we neglect the
inertial term and obtain
\begin{equation}
\frac{d\phi_i}{dt}=L+\sin(\phi_i)+\Delta^2 \phi_i.
\end{equation}
As an application, this simplified model reproduces the complex
behavior of the charged density waves (CDWs), including the
depinning transition, mode-locking, and sub-threshold
hysteresis~\cite{middleton1992asymptotic,pietronero1983nonlinear,matsukawa1984numerical},
where $\phi_i$'s describe the configuration of the charged density
wave. Our model when implemented on circle networks is equivalent to
the over-damped Frenkel-Kontorova model in the continuum limit since
$\sin (\Delta^2 x_i) \to \Delta^2 x_i$ in that limit.

Strogatz {\em et. al} also used the model below to characterize the
CDW on a ring~\cite{strogatz1988simple},
\begin{equation}
\dot{\phi_i}=L-h\sin(\phi_i-\alpha_i)+\frac{K}{n}\sum_{j=1}^n\sin(\phi_j-\phi_i)\,.
\label{eq:cdw1}
\end{equation}
The main difference between Eq.~(\ref{eq:cdw1}) and our
model Eq.~(\ref{eq:model1}) is that the coupling term in
Eq.~(\ref{eq:cdw1}) is global rather than local as in our model.

To visualize the dynamical behavior of Eq.~(\ref{eq:model1}), it is
convenient to denote the state of a node by a point on a circle
(phase space of the local dynamics) and so the state of the whole
system can be represented by a group of points on the same circle.
If $d=0$ in Eq.~(\ref{eq:model1}), then each point will move along
the circle according to the local dynamics while for $d \neq 0$ the
points will interact with their nearest neighbors. When $\lvert w
\rvert < \lvert \epsilon \rvert$, the local dynamics indicate a
unique stable fixed point while the interaction may push nodes away
from this stable equilibrium. In fact, a threshold coupling exists
which delimits regimes for stable fixed configuration and for stable
circulation along the circle, {\em i.e.}, a stable nonlinear wave on
the network. The simplicity of the model enables a detailed analysis
of this oscillatory solutions as shown in next section.


\section{Spatiotemporal patterns}
\label{sect:patt} Let's consider the equation below
\begin{equation}
\dot{x_i}=w+d\sin(x_{i+1}+x_{i-1}-2x_{i}) \,,x_i\in \mathbb{R} \label{eq:x-simp}
\end{equation}
which is the special case $\epsilon=0$ of Eq.~(\ref{eq:model1}). In
this simplified equation, a two-parameter continuous symmetry group
comes to existence: the equation is invariant under $x_m \to
x_m+c_1+mc_2\,,\forall c_1\,,c_2\in \mathbb{R}$. Of course, the
boundary conditions should be satisfied under this transformation.
For the periodic boundary condition, $c_2$ can only take discrete
values. With the notation
\begin{equation*}
  \eta_i=
 \begin{cases}
 x_{i+1}-x_{i},\quad & i\in \{1,2,...,n-1\}\\
 x_{1}-x_n,\quad & i=n\,,\\
 \end{cases}
 \end{equation*}
Eq.~(\ref{eq:x-simp}) becomes
\begin{equation}
\frac{d\eta_i}{dt}=d\sin(\eta_{i+1}-\eta_{i})-d\sin(\eta_i-\eta_{i-1})\,,
\label{eq:eta-equation}
\end{equation}
with the constraints $\sum_i^n \eta_i=2k\pi\,,
\eta_{n+1}=\eta_1+2p\pi\,, \; k,p \in \mathbb{Z}$ due to
periodicity.

\subsection{General solutions }
\begin{figure}[ht]
\centering
\includegraphics[width=13cm]{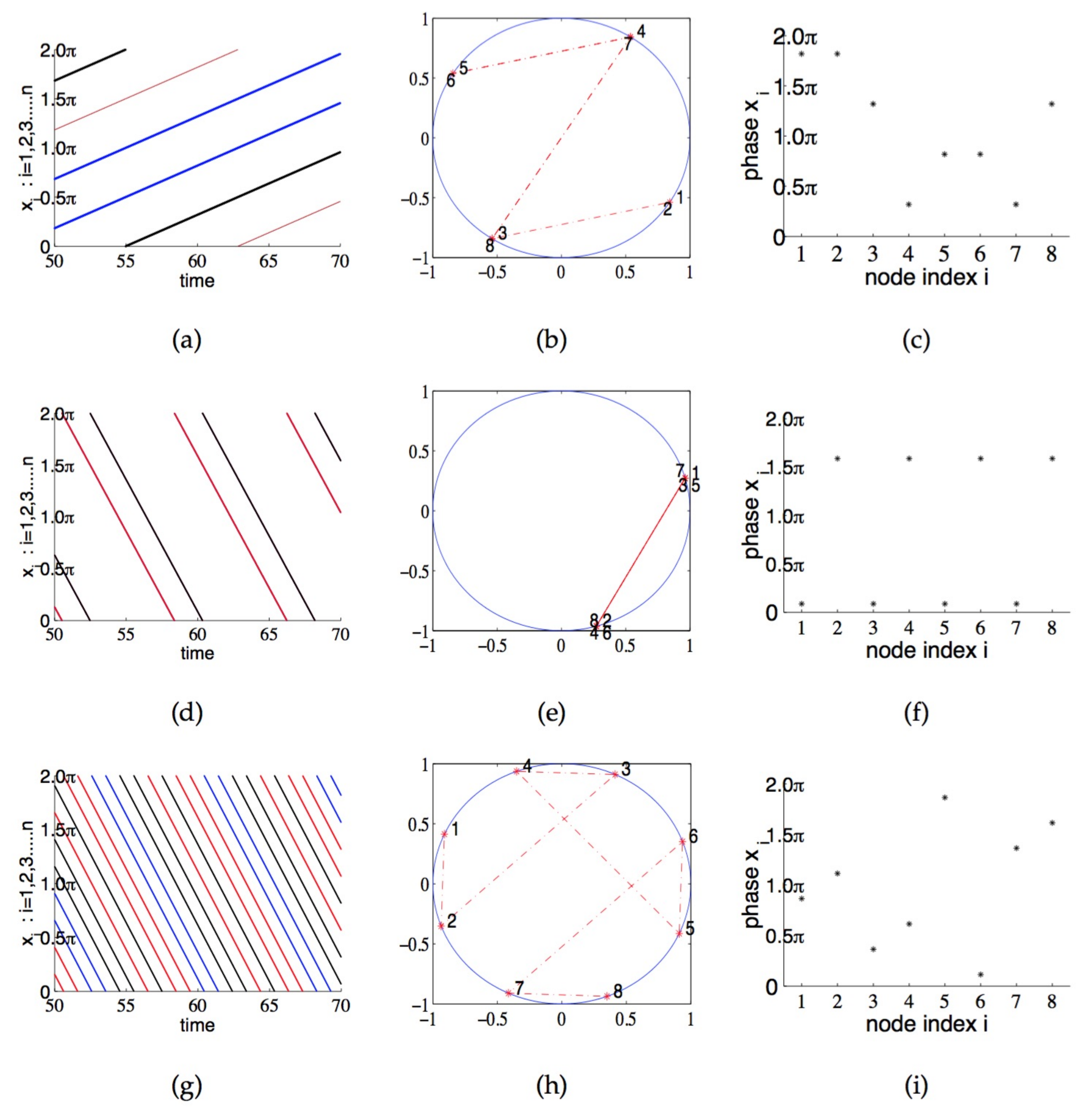}
\caption{Some interesting unstable waves on the circle network with
$n=8$. (a) An unstable wave on the circle network, corresponding to
Eq.~(\ref{eq:other1}) with $w=0.8,\,\epsilon=0,\,d=1$. (b) The phase
space configuration of a snapshot of the unstable wave in (a). (c)
The corresponding $x_i$ value on each node in (b). The plots (d),
(e), (f) and (g), (h), (i) correspond to two other unstable waves on
the same circle network. }
 \label{fig:other1}
\end{figure}

 Any stationary solution for Eq.~(\ref{eq:eta-equation})
satisfies
\begin{equation}
\,\, \eta_{i+1}-\eta_i = \beta \mbox{ or } \pi-\beta
\,,\,\forall i\,,\label{eq:gen-sol}
\end{equation}
where $\beta$ is a constant chosen to satisfy the periodic boundary
condition. So, the structure of the general solution is rather
complex even for the simplified equation. If there exist $j$'s such
that both choices of Eq.~(\ref{eq:gen-sol}) are made, in the
continuum limit, $\beta \to 0\,,\pi-\beta \to \pi$, the resulting
solution does not correspond to physical reality, which hence only
appears when the interacting units are discrete. Below is a simple
case with alternating choice of the two values
\[
\eta_{2i}-\eta_{2i-1}=\beta \,, \eta_{2i+1}-\eta_{2i}=\pi-\beta
\,,\,\,i\in 1,2,..,\mbox{ for }n=4m\,,m\in\mathbb{N} \,.
\]
We then have
\[
\eta_2=\eta_1+\beta,\,\eta_3=\eta_1+\pi,\,\eta_4=\eta_1+\pi+\beta,\mbox{
and }\eta_{i+4}=\eta_i+2\pi\,,
\]
where $\beta=k\pi/m$ to satisfy the periodicity condition when we
take $\eta_1=0$ for simplicity. Therefore,
\begin{equation}
x_2=x_1,\,x_3=x_1+\beta,\,x_4=x_1+\pi+\beta \mbox{ and }
x_i=x_{i-4}+2\pi+2\beta\,,\mbox{ for }i>4 \,, \label{eq:other1}
\end{equation}
which is a rather complex wave on the circle network. Possible
unstable nonlinear waves of this type are abundant, as illustrated
in Fig.~\ref{fig:other1}. For all  solutions in Fig.~\ref{fig:other1}, they are checked numerically to be unstable. However, a  proof for the stability of a general solution does not seem to be easy.

In order to describe waves on a circle network, a triple-plot set is
used throughout this paper, among which the first plot depicts the
time course for each node, the second plot takes a snapshot of this
dynamic wave and marks the profile in the phase space, and the third
one displays the state of each node in the second plot. This
protocol is demonstrated in Fig.~\ref{fig:other1}.

A more interesting case satisfies $\eta_{i+1}-\eta_i=\beta$
invariably such that
\begin{equation}
\eta_i=\eta_1+(i-1)\beta \,. \label{eq:eta-solve}
\end{equation}
But with the periodicity condition following
Eq.~(\ref{eq:eta-equation}), we have the constraint
\begin{equation}
\beta=\frac{2p\pi}{n}\,,\;\eta_1=\frac{2k\pi-(n-1)p\pi}{n}
\,.\label{eq:constr}
\end{equation}
The solution for the phase variable $x_i$ is
\begin{equation}
x_i=(\omega+d\sin\beta)t+(i-1)\eta_1+\frac{(i-1)(i-2)}{2}\beta+x_{00}
\,, \label{eq:eta-sol}
\end{equation}
where $x_{00}$ is some arbitrary initial phase. The stability of this solution is discussed in Appendix~\ref{app:stability}. In section
\ref{sect:bneq0}, we will discuss several special cases for this
general solution when $\beta \neq 0$, which are stable for $d\cos\beta>0$ according to Appendix~\ref{app:stability} and referred
to as special solutions or special waves in the current paper. 

When going to the continuum limit, only the solutions with
$\beta=0\,(p=0)$ makes physical sense so that the phase separations
become equal, which is referred to as the regular solution or
regular wave and will be further investigated in detail later. Under
this condition Eq.~(\ref{eq:eta-sol}) is much simplified:
\begin{equation}
x_i=\omega t+\frac{2(i-1)k\pi}{n}+x_{00} \,. \label{eq:eta-sol2}
\end{equation}
Note that the natural frequency is restored. According to Appendix~\ref{app:stability}, regular waves are stable. As we will see, the
regular wave will persist even for $\epsilon \neq 0$, where the
phase-space separations between $x_i$'s would be non-constant but
their temporal separations remain constant.

In this section, possible solutions of the model
Eq.~(\ref{eq:model1}) are classified or discussed briefly. Among
stable solutions, we identified the regular and the special waves.
Furthermore, through extensive numerical experiments we found that in
the region where nontrivial asymptotic solutions exist, basin of attraction for
regular waves or static solutions covers a large portion of the
phase space. In another word, these types of solutions most likely appear when 
starting from an initial condition chosen randomly, which 
greatly facilitates exploration of the phase space
orbit structure.
\subsection{The universality of regular nonlinear waves}
\label{sect:univ}

       \begin{figure}[ht]
       \centering
  \includegraphics[width=13cm]{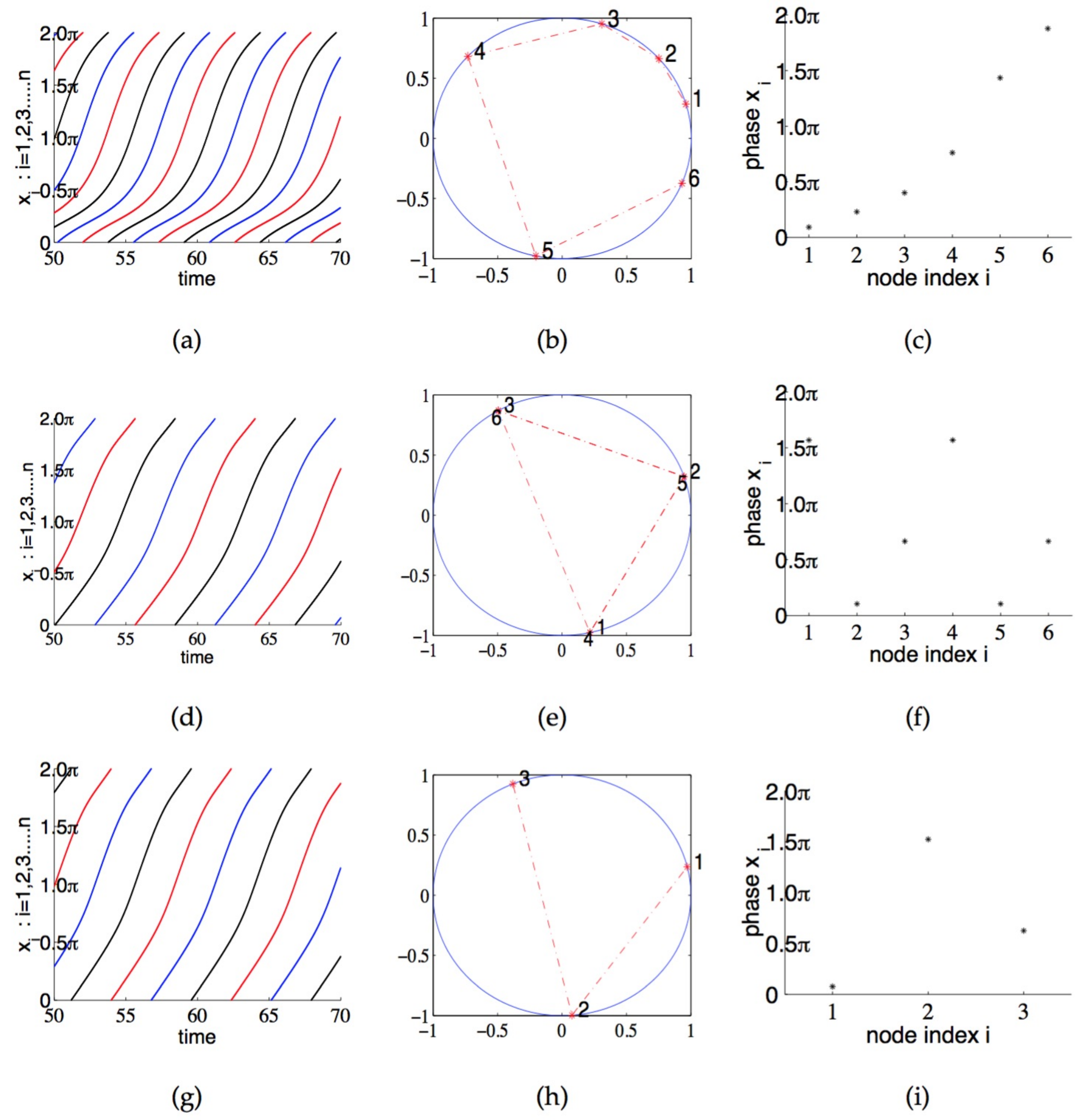}
\caption{Regular nonlinear waves on the circle network with
$w=0.8,\,\,\epsilon=1,\,\, d=1$. See Eq.~(\ref{eq:regular}).
$\langle$(a), (b), (c)$\rangle$  A regular nonlinear wave with
$k=1,\,n=6$. $\langle$(d), (e), (f)$\rangle$ A regular nonlinear
wave with $k=2,\,n=6$. $\langle$(g), (h), (i)$\rangle$ A regular
nonlinear wave with $k=1,\,n=3$.}
               \label{fig:regularsolutions}
       \end{figure}

The regular wave survives perturbation and continues to exist even
when $\epsilon $ grows as large as $\omega$ or $d$, as shown in
Fig.~\ref{fig:regularsolutions}. In circle networks, regular
nonlinear waves are the most commonly observed ones, which are
attributed to the rotational symmetry of the system and quite
independent of the local dynamics. We will make a
perturbation analysis to a generalized equation to derive the
expression of the period and the analytical form of the regular
solution to the lowest order, thus showing the universal features of
regular nonlinear waves. 

Let's consider a generalized form of Eq.~(\ref{eq:model1})
\begin{equation}
\dot{x_i}=\omega-\epsilon g(x_i)+h(x_{i+1}+x_{i-1}-2x_i),\;x_i\in \mathbb{R}
\,,\label{eq:gen-dyn}
\end{equation}
where $g\,,h$ are both smooth $2\pi$-periodic functions with
$h(0)=0,h\rq{}(0)>0,\,h\rq{}\rq{}(0)=0,\,\text{and}\,\int_0^{2\pi}g(x)dx=0$. Note that if
$h(0)\neq 0$, then $h(0)$ can be incorporated into $\omega$.
Besides, it is assumed that $a_0=\int_0^{2\pi}g(x)dx=0$, otherwise,
the transformation
\[
g(x)\rightarrow g(x)-a_0,\,\,\,\omega\rightarrow \omega-\epsilon a_0
\]
gives the right form.

The Poincar\'{e}-Lindstedt method~\cite{poincare} is a well-known
perturbation approach for approximating periodic solutions of
ordinary differential equations. An introduction to the technique
and the justification of its usage here are given in
Appendix~\ref{app:poincare}. Below, by applying the
Poincar\'{e}-Lindstedt perturbation technique, we compute the period
and the analytic form of the regular nonlinear wave to the lowest
order of $\epsilon$.  Its stability is also discussed. For
$\epsilon\ll 1$, we assume that the time and the state variable have
the following form
\begin{eqnarray}
\tau &=& \Omega t\,,\nonumber\\
\Omega &=& \omega_0+\epsilon \omega_1+\epsilon^2 \omega_2+...\,,\nonumber\\
x_i &=& x_{i,0}+\epsilon x_{i,1}+\epsilon^2x_{i,2}+...\,.
\label{eq:series}
\end{eqnarray}
Then $x_i(\tau)$ is $2\pi$-periodic and the period for $x_i(t)$ is $2\pi/\Omega$. After
substitution of Eq.~(\ref{eq:series}) into Eq.~(\ref{eq:gen-dyn}), a
comparison of different orders of $\epsilon$ leads to
\begin{equation}
1:\,\,\,\omega_0\frac{dx_{i,0}}{d\tau}=\omega+h(x_{i+1,0}+x_{i-1,0}-2x_{i,0})
\label{eq:2-1}\,,
\end{equation}
\begin{equation}
\epsilon:\,\,
\omega_1\frac{dx_{i,0}}{d\tau}+\omega_0\frac{dx_{i,1}}{d\tau}=g(x_{i,0})
+h\rq{}(x_{i+1,0}+x_{i-1,0}-2x_{i,0})(x_{i+1,1}+x_{i-1,1}-2x_{i,1})
\label{eq:2-2}\,,
\end{equation}
\begin{equation}
\begin{split}
\epsilon^2 :\,\, \omega_0\frac{dx_{i,2}}{d\tau}+\omega_1\frac{dx_{i,1}}{d\tau}
+\omega_2\frac{dx_{i,0}}{d\tau}&=x_{i,1}g\rq{}(x_{i,0})+(x_{i+1,2}
+x_{i-1,2}-2x_{i,2})h\rq{}\rq{}(x_{i+1,0}+x_{i-1,0}-2x_{i,0}) \\
&
+\frac{1}{2}h\rq{}\rq{}(x_{i+1,0}+x_{i-1,0}-2x_{i,0})(x_{i+1,1}+x_{i-1,1}-2x_{i,1})^2\,.
\end{split}
 \label{eq:2-3}
\end{equation}
With the definition $ \eta_i=x_{i+1,0}-x_{i,0}$, Eq.~(\ref{eq:2-1})
becomes
\[
\omega_0\frac{d\eta_i}{d\tau}=h(\eta_{i+1}-\eta_{i})-h(\eta_i-\eta_{i-1})\,.
\]
The regular nonlinear wave takes the form $\eta_i=-2k\pi/n,\,
k=0,1,2,3,...n-1$. Then $\omega_0=\omega$ and
\[
x_{i,0}=\tau-\frac{2(i-1)k\pi}{n}\,,
\]
which is a stable regular solution of
Eq.~(\ref{eq:2-1}). Eq.~(\ref{eq:2-2}) then becomes
\[
\omega_1+\omega
\frac{dx_{i,1}}{d\tau}=g\Big(\tau-\frac{2(i-1)k\pi}{n}\Big)+h\rq{}(0)(x_{i+1,1}
+x_{i-1,1}-2x_{i,1}) \,,
\]
which, upon substitution of the Fourier expansion for
$g(\tau-\frac{2(i-1)k\pi}{n})$, gives
\begin{equation}
\begin{split}
\omega_1+\omega\frac{dx_{i,1}}{d\tau}&=\sum_{m=1}^\infty \Big(a_m\sin\big(m(\tau-\frac{2(i-1)k\pi}{n})\big)+b_m\cos(m(\tau-\frac{2(i-1)k\pi}{n}))\Big)\\
                        &+h\rq{}(0)(x_{i+1,1}+x_{i-1,1}-2x_{i,1})\,,
\end{split}
\label{eq:2-4}
\end{equation}
where $a_m\,,b_m$ are the Fourier coefficients for $g(x)$. Let
$v_0=\sum_{i=1}^n x_{i,1}$ and a summation of the above equation
over $i$ gives
\begin{equation*}
\begin{split}
n\omega_1+\omega\frac{dv_0}{d\tau}&=\sum_{m=1}^\infty \left\{a_m\sum_{i=1}^n\Big(\sin\big(m(\tau-\frac{2(i-1)k\pi}{n})\big)\Big)+b_m\sum_{i=1}^n\Big(\cos\big(m(\tau-\frac{2(i-1)k\pi}{n})\big)\Big)\right\}\\
                              &\,\,\,\,+h\rq{}(0)\sum_{i=1}^n(x_{i+1,1}+x_{i-1,1}-2x_{i,1})\\
                              &=0+0+0\\
                              &=0\,.
\end{split}
\end{equation*}
To avoid secular terms in $v_0$, $\omega_1=0$ is taken, thus obtaining the periodicity condition
\begin{equation}
\begin{split}
\omega\frac{dx_{i,1}}{d\tau}&=\sum_{m=1}^\infty \Big(a_m\sin(m(\tau-\frac{2(i-1)k\pi}{n}))+b_m\cos(m(\tau-\frac{2(i-1)k\pi}{n}))\Big)\\
                        &+h\rq{}(0)(x_{i+1,1}+x_{i-1,1}-2x_{i,1})\,.
\end{split}
\label{eq:periodic-condition}
\end{equation}

   Note that
Eq.~(\ref{eq:periodic-condition}) is a linear differential equation with the
driving term being a superposition of trigonometric functions.
Hence, the response is a superposition of solutions for the
component-wise equation,
\begin{align*}
\dot{f}_{i}=\frac{a_m}{\omega}\sin(m(\tau-\frac{(i-1)2k\pi}{n}))+\frac{h\rq{}(0)}{\omega}(f_{i+1}+f_{i-1}-2f_{i})\,,\\
\dot{q}_{i}=\frac{b_m}{\omega}\cos(m(\tau-\frac{(i-1)2k\pi}{n}))+\frac{h\rq{}(0)}{\omega}(q_{i+1}+q_{i-1}-2q_{i})\,.
\end{align*}
According to Eq.~(\ref{eq:gen1}) and Eq.~(\ref{eq:gen2}) in
Appendix~\ref{app:force}, the solution is
\[f_{i}=a_mA_m\sin(m(\tau-\frac{(i-1)2k\pi}{n}))+a_mB_m\cos(m(\tau-\frac{(i-1)2k\pi}{n}))\,,\]
\[q_{i}=-b_mB_m\sin(m(\tau-\frac{(i-1)2k\pi}{n}))+b_mA_m\cos(m(\tau-\frac{(i-1)2k\pi}{n}))\,.\]
where
\[
A_m=\frac{4h\rq{}(0)\sin^2(k\pi/n)}{m^2\omega^2+(4\rq{h}(0)\sin^2(mk\pi/n))^2},
B_m=-\frac{m\omega}{m^2\omega^2+(4\rq{h}(0)\sin^2(mk\pi/n))^2} \,.\]
Then
\[
x_{i,1}=\sum_{m=1}^\infty
\Big((a_mA_m-b_mB_m)\sin(m(\tau-\frac{(i-1)2k\pi}{n}))+(a_mB_m+b_mA_m)\cos(m(\tau-\frac{(i-1)2k\pi}{n}))\Big)
\,.\]

Eq.~(\ref{eq:2-3}) with substitution of $x_{i,0},\,\,x_{i,1}$ and
the assumption that $h\rq{}\rq{}(0)=0$ (so complicated nonlinear
terms disappear) gives
\begin{multline}
\omega_2+\omega\frac{dx_{i,2}}{d\tau}=h\rq{}(0)(x_{i+1,2}+x_{i-1,2}-2x_{i,2})\\
+\sum_{m=1}^\infty \Big(a_mm\cos(m(\tau-\frac{2(i-1)k\pi}{n}))-b_mm\sin(m(\tau-\frac{2(i-1)k\pi}{n}))\Big)\\
\times\sum_{m=1}^\infty \Big((a_mA_m-b_mB_m)\sin(m(\tau-\frac{(i-1)2k\pi}{n}))+(a_mB_m+b_mA_m)\cos(m(\tau-\frac{(i-1)2k\pi}{n}))\Big)\,.
\end{multline}
Let $v_1=\sum_{i=1}^n x_{i,2}$, sum over $i$ and integrate from 0 to
$2\pi$ the above equation:
\begin{equation*}
\begin{split}
2\pi n\omega_2+\int_0^{2\pi}\omega\frac{dv_1}{d\tau}&=\sum_{m=1}^\infty \sum_{i=1}^{n}\Big\{\int_0^{2\pi}\Big(-b_mm(a_mA_m-b_mB_m)\sin^2(m(\tau-\frac{2(i-1)k\pi}{n}))\Big)d\tau\\
&  +\int_0^{2\pi}a_mm(a_mB_m+b_mA_m)\cos^2(m(\tau-\frac{2(i-1)k\pi}{n}))d\tau\Big\} \\
&=\sum_{m=1}^{\infty}\Big(mn\pi(a_m^2B_m+b_m^2B_m)\Big)\,.
\end{split}
\end{equation*}
To avoid secular terms in $v_1$, we take
\[
\omega_2=\frac{\sum_{m=1}^{\infty}m(a_m^2B_m+b_m^2B_m)}{2} \,.
\]

In the end, we obtain a  solution, which is stable~\cite{JKhalel},
\begin{multline}
x_i=\tau-\frac{2(i-1)k\pi}{n}+\epsilon\sum_{m=1}^\infty \Big((a_mA_m-b_mB_m)\sin(m(\tau-\frac{(i-1)2k\pi}{n}))\\
+(a_mB_m+b_mA_m)\cos(m(\tau-\frac{(i-1)2k\pi}{n}))\Big)+O(\epsilon^2)
\,,\label{eq:generalregular}
\end{multline}
with
\begin{equation}
\Omega=\omega+\frac{\sum_{m=1}^{\infty}m(a_m^2B_m+b_m^2B_m)}{2}\epsilon^2+O(\epsilon^3)\,,
\label{eq:omega}
\end{equation}
\begin{equation}
T=\frac{2\pi}{\omega}-\frac{\sum_{m=1}^{\infty}m\pi(a_m^2B_m+b_m^2B_m)}{\omega^2}\epsilon^2+O(\epsilon^3)
\,,\label{eq:general_period}
\end{equation}
where $\tau=\Omega t$ and
\begin{align*}
A_m=\frac{4h\rq{}(0)\sin^2(mk\pi/n)}{(m\omega)^2+(4h\rq{}(0)\sin^2(mk\pi/n))^2},\,\,\,\,     & a_m=\frac{1}{\pi}\int_0^{2\pi}\sin(x)g(x)dx\,, \\
B_m=-\frac{m\omega}{(m\omega)^2+(4h\rq{}(0)\sin^2(mk\pi/n))^2},
\,\,\,& b_m=\frac{1}{\pi}\int_0^{2\pi}\cos(x)g(x)dx\,.
\end{align*}
\begin{figure}[ht]
\centering
\includegraphics[width=10cm]{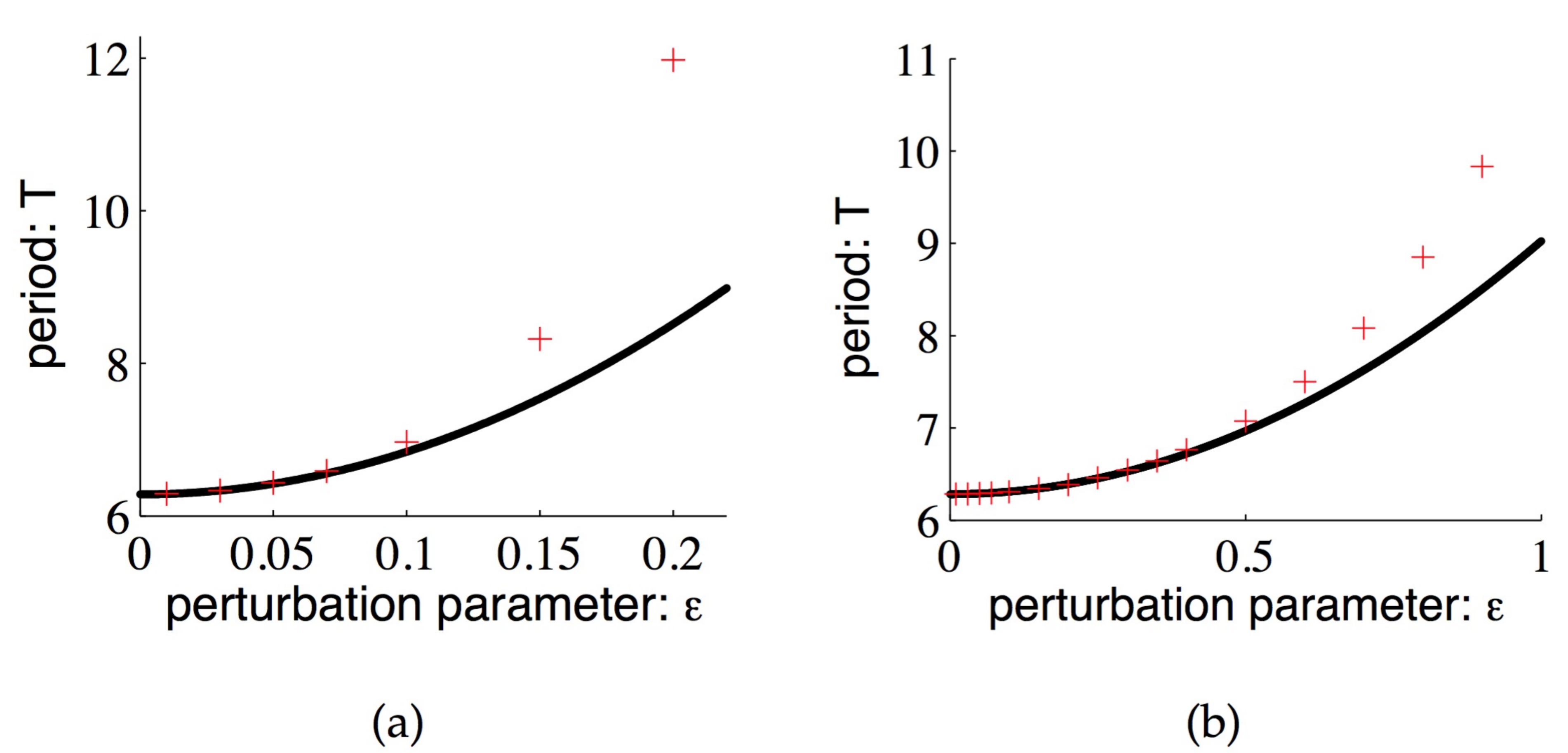}
\caption{ The dependence of  $T(\epsilon)$ on $\epsilon$ of regular
nonlinear waves. Simulation results are plotted as plus($+$), while
the analytical prediction is plotted as a solid line. (a) Results
from the system described by $ \dot{x_i}=1-\epsilon
(\sin(x_i)+2\cos(2x_i)+3\sin(3x_i)+4\cos(4x_i))+\sin(x_{i+1}+x_{i-1}-2x_i),$
with $n=10, k=1$.
 The analytical prediction is
$T=54.9444\epsilon^2+6.2832$ according to Eq.~(\ref{eq:general_period}).
 (b) Results from the system described by Eq.~(\ref{eq:model1}) with $n=10,\,\,w=1, \,\,d=1,\,\,k=1$. The
analytical prediction is $T=2.7415\epsilon^2+6.2831+O(\epsilon^3)$, based on
Eq.~(\ref{eq:regularperiod}).} \label{fig:general_period}
\end{figure}

The above perturbation analysis validated our vision that, in the
context of circle networks, regular nonlinear waves exist stably as
long as Eq.~(\ref{eq:gen-dyn}) is satisfied. Regular nonlinear waves
for different local dynamics on circle networks have been simulated
and invariably observed. In Fig.~\ref{fig:general_period}(a), we
gave an example where the period $T(\epsilon)$ {\em vs} $\epsilon$
is plotted. When $\epsilon<0.1$, the analytical results fit quite
well with the simulation results. However, when $\epsilon>0.1$ the
analytical results start to diverge, higher order terms coming into
play.

Based on the above analysis and the resulting
Eq.~(\ref{eq:generalregular}), (\ref{eq:omega}) and
(\ref{eq:general_period}), we have the following perturbation
solution for Eq.~(\ref{eq:model1}) which is a special case of
Eq.~(\ref{eq:gen-dyn})

\begin{equation}
x_i=\tau-\frac{(i-1)2k\pi}{n}+\epsilon\left(B\cos\Big(\tau-\frac{(i-1)2k\pi}{n}\Big)+A\sin\Big(\tau-\frac{(i-1)2k\pi}{n}\Big)\right)+O(\epsilon^2)
\,, \label{eq:regular}
\end{equation}
\[
\Omega=\omega-\frac{B}{2}\epsilon^2+O(\epsilon^3) \,,
\]
\begin{equation}
T=\frac{2\pi}{w}+\frac{B\pi}{w^2}\epsilon^2+O(\epsilon^3)  \,, \label{eq:regularperiod}
\end{equation}

where
\[\tau=\Omega t,\,\,A=-\frac{4d\sin^2(k\pi/n)}{w^2+16d^2\sin^4(k\pi/n)},\,\,B=\frac{w}{w^2+16d^2\sin^4(k\pi/n)}\,.\]

In Fig.~\ref{fig:regularsolutions}(a) and (d), it is easy to see
that the nonlinear wave has the property of equal time separation,
while the phase space separations between $x_i$'s are different, as
illustrated in Fig.~\ref{fig:regularsolutions}(c) and (f). In the
case of Eq.~(\ref{eq:regular}), the phase separation $\Delta t=kT/n$
for $k\neq 0$. When $k=0$, all the nodes fully synchronize and there
is no phase difference.

Thus, the perturbation analysis gives the nonlinear wave solution
and it is stable when $d>0$. On the circle network, the variable $k$
in Eq.~(\ref{eq:regular}) indicates the wavenumber for the nonlinear
wave. The corresponding wavelength $\lambda$ is $n/k$.
Regular nonlinear waves with different wavenumbers are shown in
Fig.~\ref{fig:regularsolutions}. In the phase space, wavenumbers can
be easily calculated by counting the number of circuits for which
the consecutive node $1,2,..,n$ winds around the circle. Thus, the
wavenumber is $1$ for Fig.~\ref{fig:regularsolutions}(b) and $2$ for
Fig.~\ref{fig:regularsolutions}(e). From
Eq.~(\ref{eq:regularperiod}), it is easy to see that both the period
and the frequency depend on $\epsilon$ quadratically to the lowest
order. The configuration is recurrent on the circle network in a
regular time interval $T$. The dependence of $T$ on $\epsilon$ is
computed numerically and agrees quite well with the analytical
approximation Eq.~(\ref{eq:regularperiod}) for $\epsilon<0.5$, as
shown in Fig.~\ref{fig:general_period}(b). For bigger $\epsilon$,
the actual period is larger than the analytical result, which
indicates a non-negligible role of higher order terms. It is
expected that at some finite value of $\epsilon$ depending on the
coupling $d$ and the wavenumber $k$, the regular nonlinear wave
ceases to exist and the stable solution is a fixed point which
corresponds to a uniform and stationary phase configuration.

 \begin{figure}[ht]
       \centering
  \includegraphics[width=13cm]{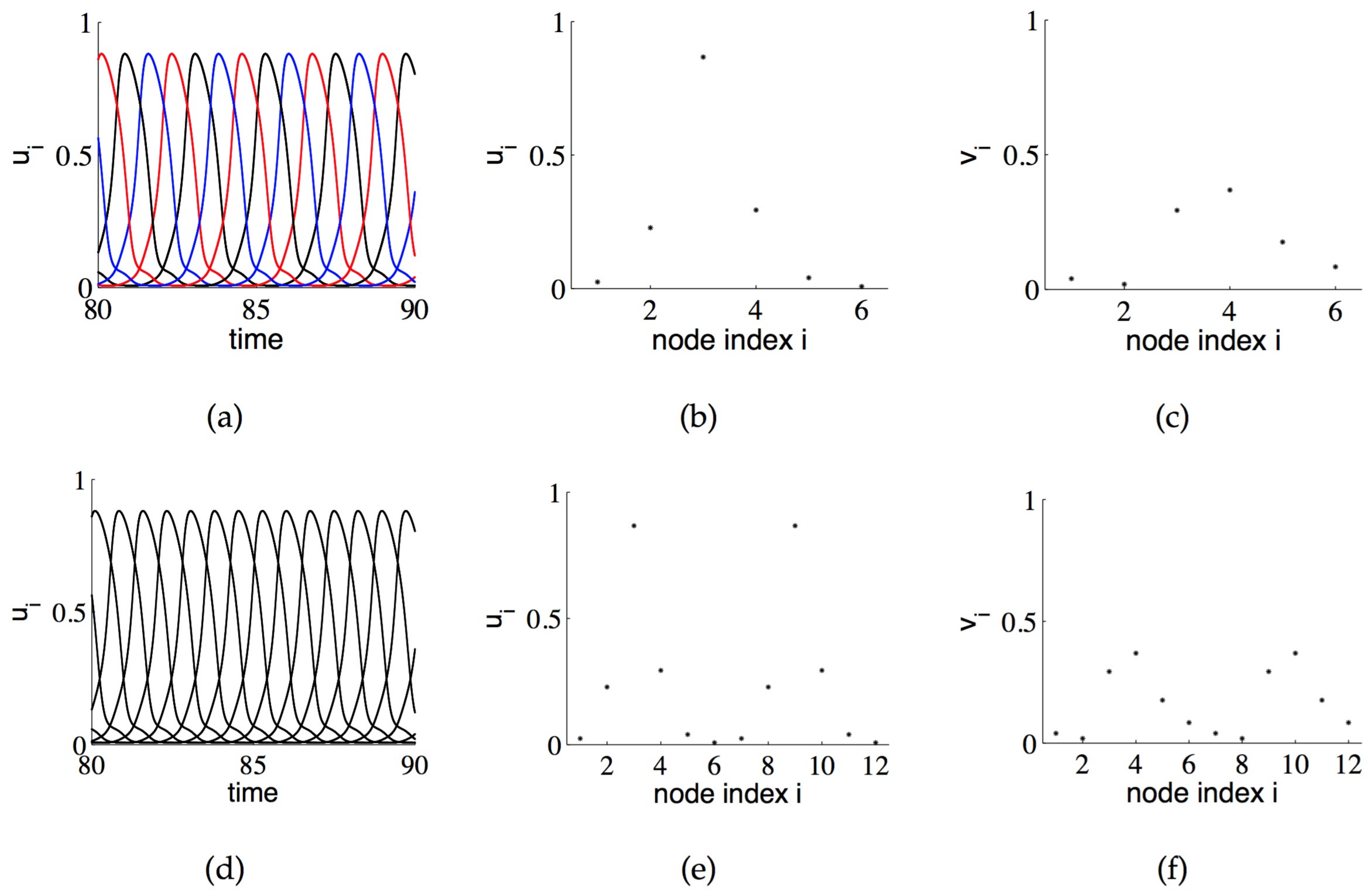}
\caption{Regular nonlinear waves on the circle network for the
revised
 B{\" a}r-Eiswirth model, with
$\epsilon=0.04,\,\,b=0.15,\,\,a=0.84,\,\,d=1$, see
Eq.~(\ref{eq:F-N-model}). In (a), (b), (c),  a regular nonlinear
wave with $k=1,\,n=6$, where (a) depicts the dynamics of $u_i(t)$,
(b) shows a snapshot of $u_i$ and (c) shows a snapshot of $v_i$. In
(d), (e), (f), a regular nonlinear wave with $k=2,\,n=12$.}
\label{fig:regularsolutions-F-N}
\end{figure}
The regular nonlinear waves in the current model are closely related
to those of the two-dimensional models, such as the B{\"
a}r-Eiswirth model~(\ref{eq:F-N-model}). Regular nonlinear waves
with wavenumbers $k=1,\,\,2$ for the B{\" a}r-Eiswirth model are
depicted in Fig.~\ref{fig:regularsolutions-F-N}. In
Fig.~\ref{fig:regularsolutions-F-N}(d), the node $i$ and node
$(i+6)(mod\,12)$ synchronize, resulting from the periodicity of the
circle network. The nonlinear wave in
Fig.~\ref{fig:regularsolutions-F-N}(d) is actually constructed from
the one in Fig.~\ref{fig:regularsolutions-F-N}(a). In simulation,
regular nonlinear waves for $n<6$ have not been found. By comparing
Fig.~\ref{fig:regularsolutions} and
Fig.~\ref{fig:regularsolutions-F-N}, we see that regular nonlinear
waves are simple yet universal for circle networks, which are
significant to the analysis of pattern formation in complex
networks.

 A key concept from the Frenkel-Kontorova model is the phase
gradient across the system, \emph{i.e.}, $\Delta x_i=x_i-x_{i-1}$.
The analysis in this work mainly addresses the case of a constant
phase gradient, \emph{i.e.}, regular solutions for $\epsilon \to 0$.
Given an arbitrary initial condition, regular nonlinear waves are
more likely to be selected according to our simulation. This can be
partly explained in terms of the soliton which is a fundamental
solution for the Frenkel-Kontorova model. More complicated wave
patterns can be constructed with multiple solitons. Presumably, due
to the repulsive interaction between the solitons, the phase
gradient tends to become uniform when the lattice pinning
effect~\cite{braun2004frenkel} vanishes, \emph{i.e.}, $\epsilon \to
0 $, thus resulting in regular nonlinear waves.

\subsection{Stable solutions with $\beta \neq 0$}
\label{sect:bneq0}

Stable solutions for $\beta \neq 0$ are also observed and only
appear in the discrete case, as illustrated in
Fig.~\ref{fig:special1}. This special type of solutions, if viewed
as nonlinear waves, has nearly all nodes synchronized in pairs,
leaving one (when $n$ is odd) or two (when $n$ is even) nodes moving
alone.

In the general solution (\ref{eq:eta-sol}), when n is odd, we may
take $k=(n+1)/2$ in the constraint Eq.~(\ref{eq:constr}) such that
$\eta_1=\beta$. In this case,
\begin{equation}
x_i=x_{n+1-i}, i\in \{1,2,3,..,\frac{n-1}{2}\}\,.\\
\label{eq:specialodd}
\end{equation}

When $n$ is even, we may take $k=n/2$ in Eq.~(\ref{eq:constr}) so
that $\eta_1=\beta/2$. In this case,
\begin{equation}
x_i =x_{n+2-i}, i\in \{2,3,..,\frac{n}{2}\} \,.\label{eq:specialeven}
\end{equation}

The discussion above is also applicable to the circumstance with
$\epsilon > 0$. Fig.~\ref{fig:special1} demonstrates two special
typical nonlinear waves with $\epsilon=0.05$.  In
Fig.~\ref{fig:special1}(b), we see that the node pairs
$(1,7),\,(2,6),\, (3,5)$ synchronize while node $4$ is left alone,
in accordance to Eq.~(\ref{eq:specialodd}). In
Fig.~\ref{fig:special1}(e), the node pairs
$(2,10),\,(3,9),\,(4,8),\,(5,7)$ synchronize while node $1$ and $6$
are left alone, corresponding to Eq.~(\ref{eq:specialeven}).
Comparison of configurations has been made between special nonlinear
waves computed numerically with $\epsilon>0$ (red stars) and their
counterparts with $\epsilon=0$ by Eq.~(\ref{eq:eta-sol}) (blue
stars), as illustrated in Fig.~\ref{fig:special1}(b) and (e). It
seems that the analytical solution with $\epsilon=0$ is a good
approximation for small $\epsilon$.

When $\epsilon=0$, we have
\begin{equation}\label{eq:special-period}
T=\frac{2\pi}{w+d\sin(\beta)}\,,
\end{equation}
where $\beta=2p\pi/n$. In all special nonlinear waves, the ones with
smaller $\lvert p \rvert$ are more likely to be observed. Compared
with the period of the counterpart regular nonlinear wave with the
same $\omega,\epsilon,d$, negative $p$'s lead to longer periods and
positive $p$'s to shorter ones. To show how well
Eq.~({\ref{eq:special-period}) approximates the period of the
special nonlinear wave with $\epsilon \neq 0$, a numerical result is
displayed in Fig.~\ref{fig:special2}. The analytical expression
gives good prediction when $\epsilon$ is relatively small. The
relative error is less than $2\%$ when $\epsilon \le 0.2$. When
$\epsilon$ increases over a critical value $\epsilon_c$, the special
nonlinear wave becomes unstable. In each case displayed in
Fig.~\ref{fig:special2}, the result with the largest $\epsilon$
({\em i.e.}, equal to $\epsilon_c$) corresponds to the special
nonlinear wave that is on the brim of being unstable. As the
simulation shows, $\epsilon_c$ depends on $d,\,w,\,n,\,p$. Stable
special nonlinear waves do not exist whenever $\epsilon>\omega$.
Therefore, in this case, we may focus on regular nonlinear waves,
which are easier to analyze.

In Fig.~\ref{fig:special2}(b) and (d) for $p=-1$, the dependence of
the period $T$ shows a non-monotonic behavior on the size of
perturbation: it increases when $\epsilon$ is small but decreases
after passing a maximum, which is very different from that of the
regular solution. The detailed dependence of $T(\epsilon)$ on
$\epsilon$ and its physical implication remain to be explored.

Special nonlinear waves seem universal in circle networks since they
are easily observed in the general 1-dimensional model
(\ref{eq:master}) with different local dynamics. In
Fig.~\ref{fig:general_period}(b), an example is presented where
special nonlinear waves are observed for $\epsilon \le 0.22$.
The special solution  corresponds to a finite curvature in
the phase field, \emph{i.e.}, variable phase gradient in
Frenkel-Kontorova model~\cite{braun2004frenkel}.

        \begin{figure}[ht]
       \centering
    \includegraphics[width=13cm]{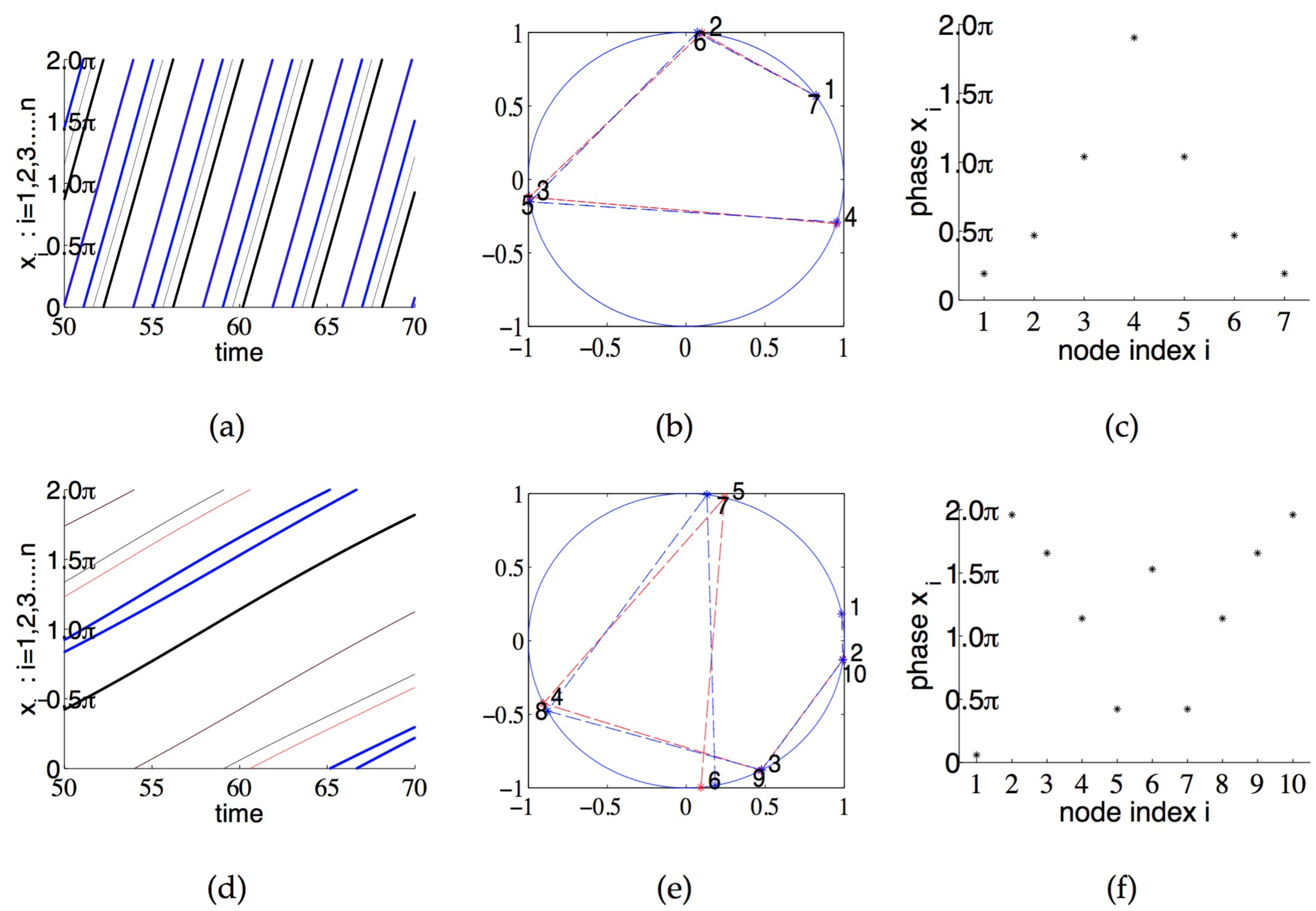}
\caption{Typical stable special nonlinear waves. (a), (b), (c) plot
a special nonlinear wave on the circle network, with
$n=7,\,w=0.8,\,\epsilon=0.05,\,d=1,\,p=1$ for
Eq.~(\ref{eq:specialodd}) and Eq.~(\ref{eq:special-period}). (d),
(e), (f) plot a special nonlinear wave on the circle network, with
$n=10,\,w=0.8,\,\epsilon=0.05,\,d=1,\,p=-1$ for
Eq.~(\ref{eq:specialeven}) and Eq.~(\ref{eq:special-period}). Blue
stars in (b) and (e) indicate the corresponding analytical solution
with $\epsilon=0$ in Eq.~(\ref{eq:eta-sol}). }
   \label{fig:special1}
       \end{figure}

      \begin{figure}[ht]
       \centering
  \includegraphics[width=10cm]{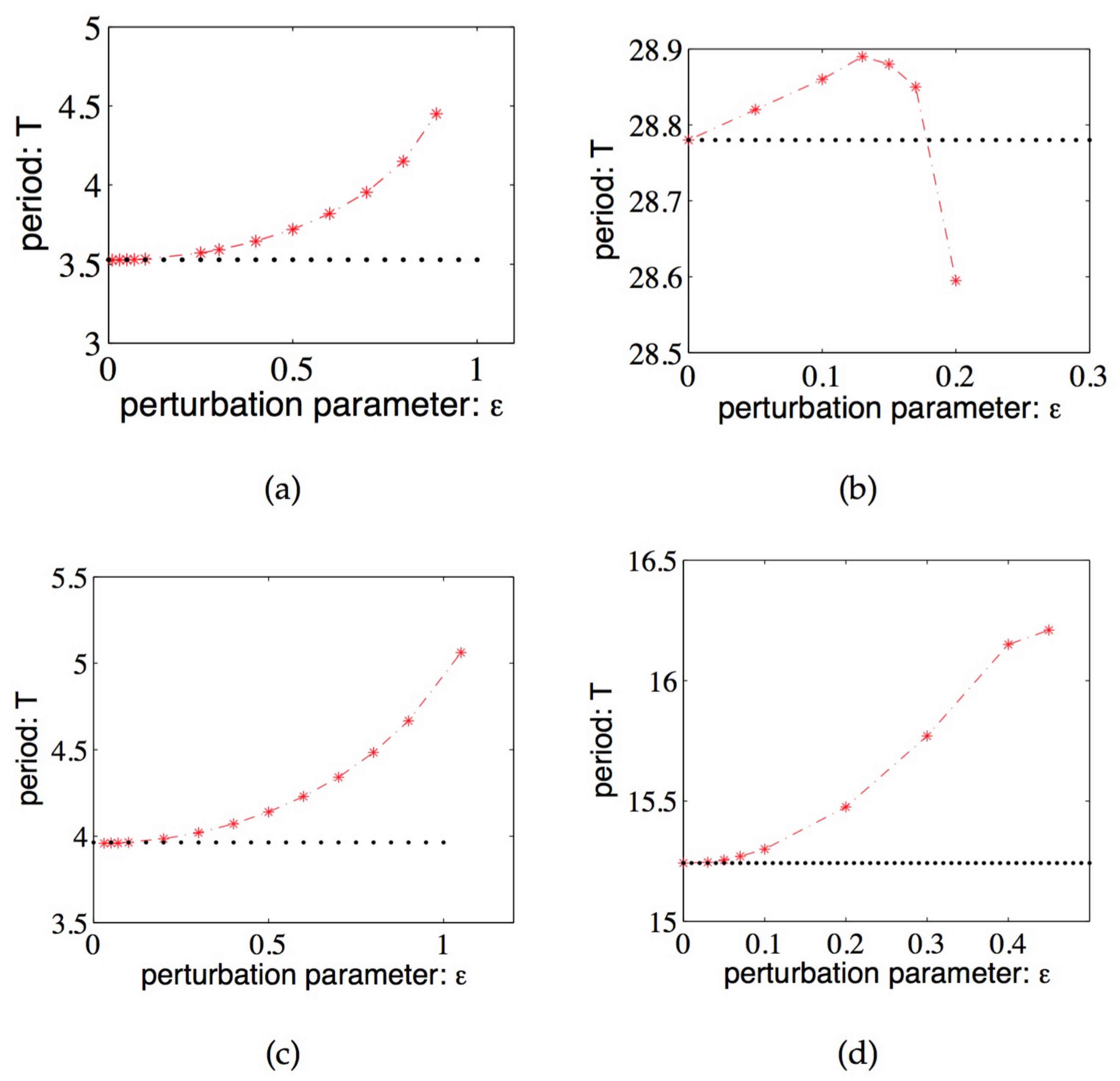}
\caption{The dependence of the period on $\epsilon$ for different
stable special nonlinear waves. Simulation results are plotted as
red stars (with dashed line), along with the analytic result for $\epsilon=0$ serving
as the baseline, which are plotted as black dots. (a) The parameter
values are $n=7,\,w=0.8,\,d=1,\,p=1$ in Eq.~(\ref{eq:specialodd})
and Eq.~(\ref{eq:special-period}). (b) The same with (a) except that
$p=-1$. (c)  $n=10,\,w=0.8,\,d=1,\,p=1$ in
Eq.~(\ref{eq:specialeven}) and Eq.~(\ref{eq:special-period}). (d)
The same with (c) but $p=-1$. } \label{fig:special2}
\end{figure}

\section{Circulating pulse and regular nonlinear wave}
\label{sect:corr}

\begin{figure}[ht]
\centering
\includegraphics[width=10cm]{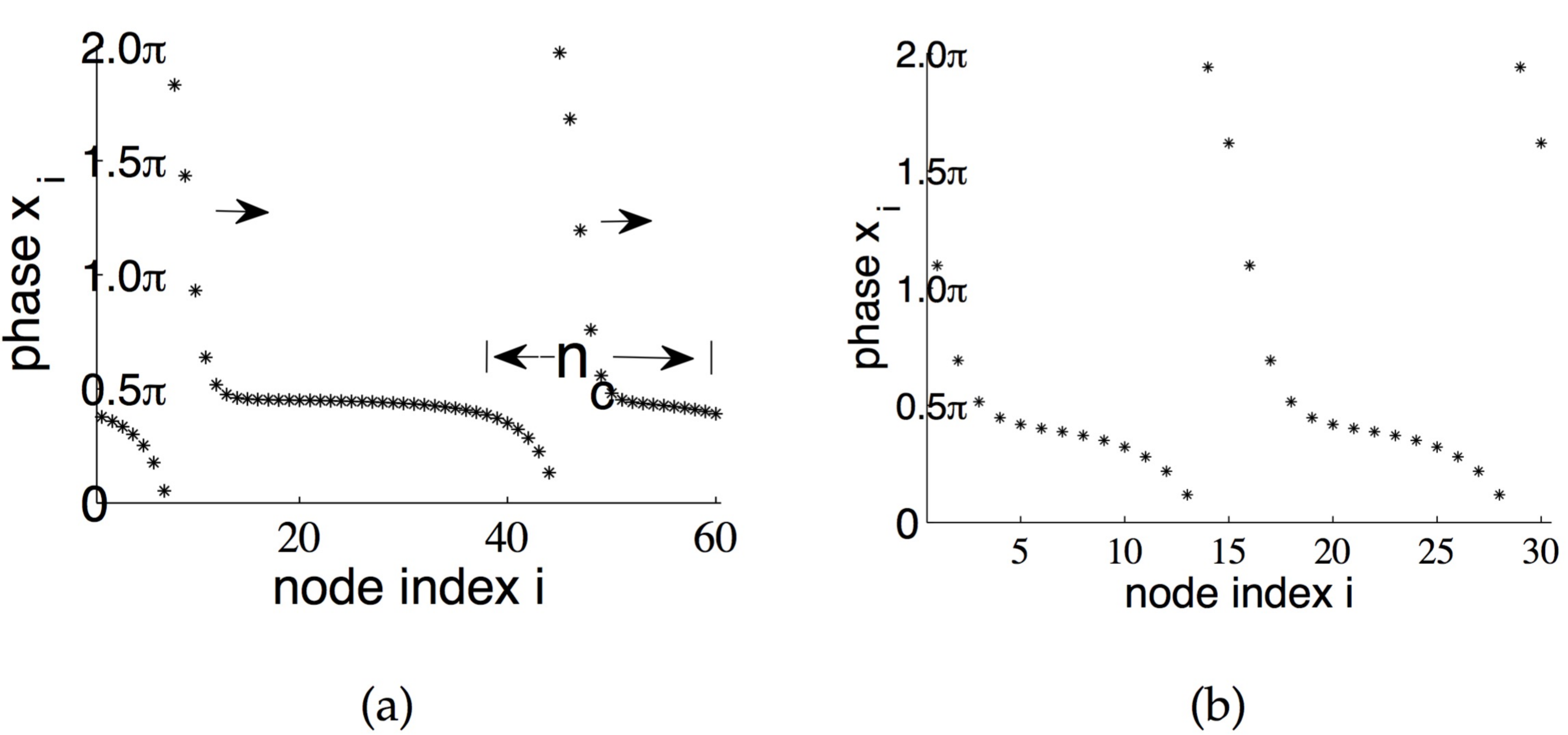}
\caption{(a) Two independent circulating pulses on a circle network
with $n=60,\,w=0.99,\,\epsilon=1,\,d=1$. The moving direction is
indicated by the arrow. $n_c$ is the number of nodes covered by one
pulse. (b) A regular nonlinear wave with
$k=2,\,n=30,\,w=0.99,\,\epsilon=1,\,d=1$.} \label{fig:pulse2}
\end{figure}
Our model is closely related to the excitable media on a circle. In
previous studies, a circulating pulse was rendered unstable and
changed to a steady solution upon decreasing the circumference of
the ring~\cite{instability1993,frame1988oscillations}. In our
discrete model, pulses would first change to regular nonlinear waves
with decreasing $n$. The difference between circulating pulses and
regular nonlinear waves are explained in Fig.~\ref{fig:pulse2}. In
Fig.~\ref{fig:pulse2}(a), the two pulses do not constitute a regular
nonlinear wave, since the distance between them may be adjusted
freely as long as it is larger than the critical number $n_c$, where $n_c$ is 
the number of nodes spanned by a single pulse , thus contradicting the uniqueness of the regular nonlinear
wave with $k=2$. Note that the node indices corresponding to peaks
are $i_{peak1}=8,i_{peak2}=45$ and thus $i_{peak2}-i_{peak1}\neq
n/2=30$, while $i_{peak2}-i_{peak1}= n/2$ for a regular nonlinear
wave, as depicted in Fig.~\ref{fig:pulse2}(b). When $n$ decreases to
30, the two pulses strongly interact with each other and a regular
nonlinear wave with $k=2$ emerges. So a circulating pulse can be
viewed as a local structure, which spreads over $n_c$ nodes in the
circle network, as indicated in Fig.~\ref{fig:pulse2}(a). Two pulses
will change to a regular nonlinear wave if they overlap. This
phenomenon is characteristic of the discreteness of our model on a
circle network.

The circulating pulse is similar to the soliton solution of
the Frenkel-Kontorova model. The stability of circulating pulse is
due to the excitable property of individual nodes while the
stability of soliton is also well explained in the Frenkel-Kontorova
model in terms of the lattice pinning
effect~\cite{braun2004frenkel}, \emph{i.e.}, potential well created
by the discreteness of lattice. Our model describe the lattice
pinning effect and the excitability property in an equivalent and
unified way.

 \begin{figure}[ht]
 \centering
\includegraphics[width=10cm]{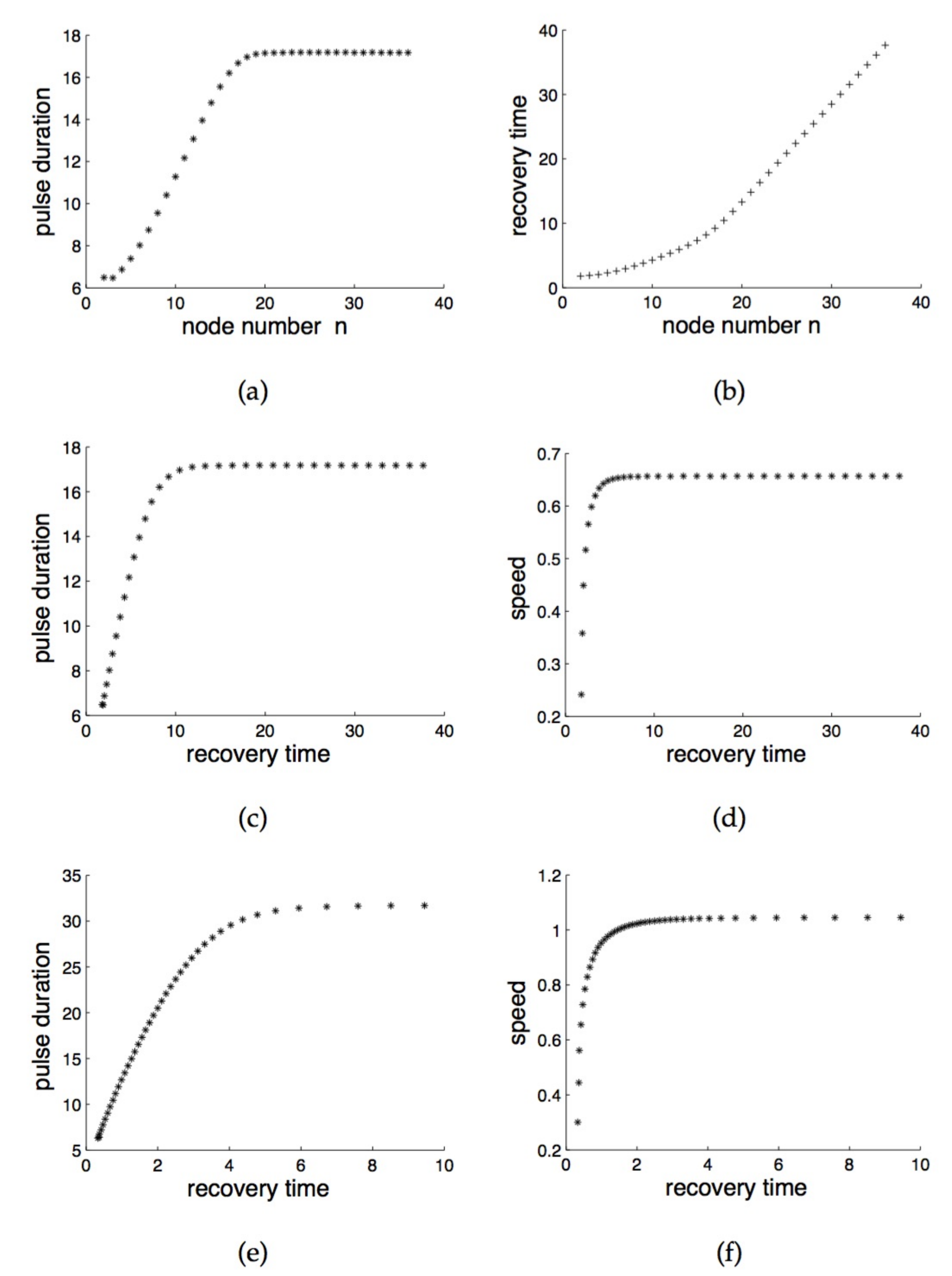}
\caption{Interesting characterizations of the circulating pulse in a
circle network. $d=1,\,\epsilon=1$ in all plots. (a) The pulse
duration $a$ as a function of the total node number $n$, with
$\omega=0.8$. (b) The recovery time $t_r(n)$, with $\omega=0.8$. (c)
The restitution curve $a(t_r)$ with $\omega=0.8$. (d) The dispersion
curve $c(t_r)$ with $\omega=0.8$. (e) The restitution curve $a(t_r)$
with $\omega=0.99$. (f) The dispersion curve $c(t_r)$ with
$\omega=0.99$. }
 \label{fig:pulse-recovery-duration}
 \end{figure}

The recovery time and the pulse duration are two important physical
observables in the evolution of excitable node dynamics and have
been well characterized in the literature~\cite{instability1993}. In
our case, the recovery time $t_r$ is the time spent in overcoming
the barrier between two fixed points defined by the local dynamics
({\em i.e.}, $x_{fix1}=\arcsin(w),\,x_{fix2}=\pi-\arcsin(w)$).
$t_r=t_2-t_1$, where $x(t_1)=x_{fix1}, \,\,x(t_2)=x_{fix2}$. The
pulse duration $a=T-t_r$ is the time spent in the long excursion
away from the fixed points. Due to the rotational symmetry of our
model, $t_r$ and $a$ are independent of node position in the circle
network, but strongly dependent on the circumference $n$. A careful
simulation is undertaken to determine the curves
$t_r(n),\,a(n),\,a(t_r),\,c(t_r)$, shown in
Fig.~\ref{fig:pulse-recovery-duration}, where $c$ is the speed of
the circulating wave defined by $c=n/(t_r(n)+a(n))$. The curve
$a(t_r)$ is termed the \emph{restitution curve} and $c(t_r)$ the
\emph{dispersion curve}. These curves are important in that they
help us understand relevant and universal properties for circulating
pulses in a circle network~\cite{instability1993}.

In Fig.~\ref{fig:pulse-recovery-duration}(a) and (b), there is a
turning point in $a(n)$ and $t_r(n)$ around $n=18$, which signifies
a transition from a regular nonlinear wave to a circulating pulse.
The saturation in $a(n)$ and the constant slope for $t_r(n)$ when
$n>18$ is characteristic of a circulating pulse, simply because the
pulse does not spread over all nodes and moves at a constant speed
along the circle network for $n$ large enough. Thus the critical
number $n_c$ of nodes covered by a pulse can be defined as $n$ at
the transition point. Geometrically, on the phase circle, an
increasing number of nodes are moving at the recovery stage with a
decreasing speed and thus raised the recovery time. The density of
the nodes on the excursion contour remains constant for large $n$,
leading to a constant duration time. The \emph{restitution curve}
$a(t_r)$ and the \emph{dispersion curve} $c(t_r)$ are plotted in
Fig.~\ref{fig:pulse-recovery-duration} with different $\omega$'s
which look qualitatively similar. The difference originates from
disparate distances between fixed points of the local dynamics. The
saturation of $a(t_r)$ and $c(t_r)$ are also associated with the
transition from a regular nonlinear wave to a pulse. Similar
saturation behavior was observed in a one-dimensional ring of
excitable media with very sophisticated description of motion
dynamics~\cite{instability1993}. The physical observables
$t_r(n),\,a(n),\,a(t_r),\,c(t_r)$ used here are universal quantities
for characterizing circulating pulses in a circle network.

\section{Summary}
\label{sect:sum} Previous works~\cite{qian2010structure} suggest
that dynamics of coupled excitable nodes on a complex network may
have a very simple yet universal structure, which includes
self-sustained oscillation on the circle sub-network and attached
branches driven by the center oscillation. This article focuses on
discussion on possible solutions for a new type of equation, which
is simple enough for analytic computation yet captures the essence
of nonlinear wave generation and propagation on networks with
excitable nodes. This new model can be viewed as a most direct
extension of the Kuramoto model to treat the excitable dynamics, the
understanding of which will help us study possible behaviors of
other models with excitable dynamics on complex networks due to its
universality.

In this paper, we carried out a quite thorough study of the new
model on circle networks and reveals certain universality of regular
solutions, which is quite independent of the local dynamics and the
attached branches being driven, and closely related to the Laplacian
coupling of neighboring nodes. Although there are numerous solutions
for this system, in terms of stability and basin of attraction,
regular nonlinear waves and, in the case of large number of nodes,
circulating pulses are the most important solutions, which was
confirmed by simulation. The period $T$ for regular nonlinear waves
is computed analytically which agrees well with the numerical result
up to $\epsilon \sim 0.5\omega$. An analytic form of the regular
nonlinear wave is also obtained to the first order. A new type of
solution, the special nonlinear wave, is also studied and compared
to the regular solution. It is stable under certain parameter regime
but only exists in the discrete dynamics. The properties of
circulating pulses on circle network with excitable nodes were
discussed in detail and its relevance to the regular solution is
also studied.

Our model is closely related to the Frenkel-Kontorova model.
While the classical Frenkel-Kontorova model describes the Newtonian
dynamics of a chain of classical particles, here we use the
over-damped version with a special form of coupling and use it to
study pattern formation on complex networks. The simplicity of this
model and previous understanding of the Frenkel-Kontorova model
hopefully give physical insights while still keeping the analytic
understanding within reach.

Much more work needs to be done. For example, the existence
condition and the basin of attraction for each regular solution
should be more precisely characterized. Further, we may employ the
current model to study the interaction of the circle sub-network and
the attached branches in a network of general topology, or the
interaction of real complex networks. It would be interesting to
compare wave propagation on the same network but with different
excitable dynamics.

\section*{Acknowledgements}
This research is supported by National Natural Science Foundation of
China (Grant No. 10975081).

\appendix
\section{Stability of solutions}
\label{app:stability}
Let us consider the stability of solutions for Eq.~(\ref{eq:eta-equation}). We write the perturbation solution  in this form $\eta_i^{(p)}=\eta_i+\epsilon_i$, in which $\eta_i$ is the stationary solution for Eq.~(\ref{eq:eta-equation}) and satisfies Eq.~(\ref{eq:eta-solve}). Ignoring higher order terms of $\epsilon$, a substitution of  $\eta_i^{(p)}$ into Eq.~(\ref{eq:eta-equation}) gives
  \begin{equation}
\frac{d\epsilon_i}{dt}=\alpha(\epsilon_{i+1}+\epsilon_{i-1}-2\epsilon_i)\,,
\label{eq:stability2}
\end{equation}
where $\alpha=d\cos(\beta)$. Suppose $\alpha>0$, then 
\[\frac{d\vec{\epsilon}}{d\tau}=-\alpha Q\vec{\epsilon}=A\vec{\epsilon}\,.\]
So $Q$ can be viewed as a Laplacian matrix of the circle network.
 It turns out that $Q$ is a
positive semi-definite matrix, with eigenvalues $0=\lambda\rq{}_0\le
\lambda\rq{}_1 \le...\le \lambda\rq{}_{n-1}$. So the eigenvalues
$\lambda_i$ for $A=-\alpha Q$ are all non-positive if $\alpha>0$.

More specifically, suppose
\[\epsilon_i=ae^{j\frac{2mi\pi}{n}}e^{\lambda_m\tau} ,m=0,1,...n-1\,,\]
where $j=\sqrt{-1}$. Then $\lambda_m=-4\alpha\sin^2(m\pi/n)$, which
are the eigenvalues of this system. All the eigenvectors for $m
> 0$ correspond to the stable direction and the one for the eigenvalue
$\lambda_0=0$ corresponds to the rotation of the system as a whole.
Therefore, up to a rotation, the solution is stable and the larger
$\alpha$ is, the more stable the solution will be. From the discussion above, we  conclude that $\alpha>0$ gives stable solutions while $\alpha=0$ neutrally stable solutions and $\alpha<0$ unstable solutions. 


However, the following situation
\[\exists\, m,\,\eta_{m+1}-\eta_m=\beta\,\,\text{and}\,\eta_{m}-\eta_{m-1}=\pi-\beta\]
for the stationary solution of Eq.~(\ref{eq:eta-equation}) is more difficult to analyze. This situation may possibly
though not necessarily bring instability to the system. Numerical
simulation results show that that the system would be unstable under
this condition in most cases except a few.

\section{The Poincar\'{e}-Lindstedt method}
\label{app:poincare}

 In seeking for a periodic solution, the regular
perturbation technique usually fails because the period of the new
solution is slightly different from that of the unperturbed one
which the perturbation expansion starts from. The mismatch of the
two periods usually results in secular terms which grow without
bound. The Poincar\'{e}-Lindstedt method overcomes this shortcoming
by allowing stretching or compressing of the time coordinate, thus
matching the two periods. To implement this method, we need to make
sure that the periodic solution exists and the expansion converges
at least for the small perturbation. Below, we briefly review the
theorem on the existence of periodic solutions and then verify that
our general model satisfies the existence conditions.

\subsection{The existence of periodic solutions}
Consider in $\mathbb{R}^n$ the equation
\begin{equation}
\dot{x}_i=f_i(\vec{x})+\epsilon g_i(\vec{x})\,, \label{eq:exist}
\end{equation}
where $\epsilon$ is a small parameter, $i\in\{1,2,...,n\}$,
$\vec{x}=(x_1,x_2,...,x_n)$, $f_i(\vec{x})=f_i(x_1,x_2,...,x_n)$ and
$g_i(\vec{x})=g_i(x_1,x_2,...,x_n)$. We are seeking the periodic
solutions of the equation. For the unperturbed equation
\[\dot{y}_i=f_i(\vec{y})\,,\]
it is assumed that a $T_0$-periodic solution
\begin{equation}
y_i=G_i(t;\varphi_1,\varphi_2,...,\varphi_n)
\label{eq:exist-yi}
\end{equation}
exists with $\varphi_i$\rq{}s being integral constants. To get a
periodic solution for Eq.~(\ref{eq:exist}), it is convenient to make
a coordinate transformation
 \[\Omega t=\tau,\,\Omega^{-1}=\omega^{-1}+\epsilon \phi_0 (\epsilon),\]
with $\Omega$ being the frequency of the new solution and
$\omega=2\pi/T_0$.  This transformation allows us to directly
approximate the new period. After this transformation, the new
sought solution is $2\pi$-periodic in $\tau$.
A perturbation expansion of the solution for Eq.~(\ref{eq:exist})
could be written as
  \begin{equation}
  x_i=M_i(\tau;\phi_0,\phi_1,\phi_2,...,\phi_n,\epsilon).
  \label{eq:exist-xi}
  \end{equation}
Denote
\[H_i(\vec{\phi},\epsilon)=M_i(2\pi;\phi_0,\phi_1,\phi_2,...,\phi_n,\epsilon)-M_i(0;\phi_0,\phi_1,\phi_2,...,\phi_n,\epsilon),\]
where $\vec{\phi}=(\phi_0,\phi_1,...,\phi_n)$. The periodicity
condition requires
\begin{equation}
H_i(\vec{\phi},\epsilon)=0\,, \label{eq:condiperio}
\end{equation}
which determines $n$ parameters from the $n$ periodicity conditions
Eq.~(\ref{eq:condiperio}) in the neighborhood of $\epsilon=0$. The
extra parameter corresponds to the time translational symmetry of
the autonomous equations. Eq.~(\ref{eq:condiperio}) is equivalent to
\begin{equation}
\text{rank}\Big(\frac{\partial \vec{H}}{\partial \vec{\phi}}\Big)=n,
\label{eq:unique}
\end{equation}
where $\vec{H}=(H_1,H_2,...,H_n)^T$ and $\frac{\partial
\vec{H}}{\partial \vec{\phi}}$ is the Jacobian matrix of $\vec{H}$.
This is also called the uniqueness condition since it determines
uniquely the new periodic solution up to a time translation.

Below we state theorems relevant to the Poincar\'{e}-Lindstedt
method. See~\cite{JKhalel}[Chapter 9 and 10] for more details.
\vspace{3ex}

\textbf{The Poincar\'{e} expansion theorem} addresses the problem of
the convergence of the usual perturbation solution of differential
equations within certain time scale. And it roughly states that if
$f_i(\vec{x})$ and $g_i(\vec{x})$ of Eq.~(\ref{eq:exist}) can be
expanded in a convergent power series with respect to $\vec{x}$,
then regular perturbation series converge in the neighborhood of
$\epsilon=0$ and the original initial condition within a time-scale
1.

\textbf{The uniqueness theorem of periodic solutions} concludes that
If the uniqueness condition Eq.~(\ref{eq:unique}) is satisfied for
Eq.~(\ref{eq:exist}), along with the requirement of the Poincar\'{e}
expansion theorem and the periodicity condition
Eq.~(\ref{eq:condiperio}), then there exists a periodic solution
which can be represented by a convergent power series in $\epsilon$
in the form of Eq.~(\ref{eq:exist-xi}) for $0\le\epsilon<\epsilon_0$
for some positive $\epsilon_0$.

\subsection{Justification}
Now we justify the application of the Poincar\'{e}-Lindstedt method
in our generalized model Eq.~(\ref{eq:gen-dyn}), which apparently
satisfies the condition of the Poincar\'{e} expansion theorem since
$g(x),\,h(x)$ are both smooth functions. Below, we check the
uniqueness and periodicity conditions.

Under the coordinate transformation
    \[\Omega t=\tau,\,\Omega^{-1}=\omega^{-1}+ \phi_0 (\epsilon)\,,\]
where $\phi_0(\epsilon)=O(\epsilon)$, Eq.~(\ref{eq:gen-dyn}) becomes
  \begin{equation}
  \frac{dx_i}{d\tau}=1+\omega \phi_0-(\frac{\epsilon}{\omega}+\epsilon \phi_0)g +(\phi_0 +\frac{1}{\omega})h.
   \label{eq:3-1}
   \end{equation}
   For the unperturbed equation
   \[ \frac{dy_i}{d\tau}=1+\frac{1}{\omega}h(y_{i+1}+y_{i-1}-2y_i),\]
   the regular solution is
   \[y_i(\tau)=\tau-\frac{2(i-1)k\pi}{n}\,.\]
We can express the regular solution for Eq.~(\ref{eq:3-1}) as
    \begin{equation}
   \begin{split}
   &x_i(\tau)=\tau-\frac{2(i-1)k\pi}{n}+\phi_i+\\
   & \int_0^\tau \Big(\omega\phi_0-(\frac{\epsilon}{\omega}+\epsilon\phi_0)g(x_i(\theta))+(\phi_0 +\frac{1}{\omega})h(x_{i+1}(\theta)+x_{i-1}(\theta)-2x_i(\theta))\Big)d\theta,
\end{split}
\end{equation}
where  $\phi_i=O(\epsilon)$ is a constant parameter and
$\phi_i|_{\epsilon=0}=0$. In accordance with the Poincar{\' e}
expansion theorem, we assume that
\begin{equation}
y_i=x_i+O(\epsilon),
\label{eq:xi-yi}
\end{equation}
then  $x_{i+1}(\theta)+x_{i-1}(\theta)-2x_i(\theta) \approx O(\epsilon)$, thus
\[h(x_{i+1}+x_{i-1}-2x_i)=h(0)+h`(0)O(\epsilon)+o(\epsilon)= O(\epsilon).\]
Then
\[\int_0^\tau \left(\omega\phi_0-(\frac{\epsilon}{\omega}+\epsilon\phi_0)g(x_i)+(\phi_0 +\frac{1}{\omega})h(x_{i+1}+x_{i-1}-2x_i)\right)d\theta=O(\epsilon),\]
which is consistent with Eq.~(\ref{eq:xi-yi}). A substitution of $y_i$
for $x_i$ at the right side of Eq.~(\ref{eq:3-1}) with the notation
$z_i=y_i+\phi_i$ gives
\begin{equation}
x_i(\tau)=z_i+  \int_0^\tau
\left((\omega\phi_0-\frac{\epsilon}{\omega}g(z_i))+\frac{h`(0)}{\omega}(z_{i+1}+z_{i-1}-2z_i)\right)d\theta+o(\epsilon).
\end{equation}

The periodicity condition is
\[x_i(2\pi)=x_i(0)+2\pi\,,\]
which, with the notation
  \[H_i=\int_0^{2\pi} \left((\omega\phi_0-\frac{\epsilon}{\omega}g(z_i))+\frac{h`(0)}{\omega}(z_{i+1}+z_{i-1}-2z_i)\right)d\theta+o(\epsilon).\]
is equivalent to
  \[H_i(\phi_0,\phi_1,...,\phi_n)=0, \,i\in \{1,2,...,n\}.\]
Below, we calculate the Jacobian matrix of $\vec{H}$ in the
neighborhood of $\epsilon=0$ and $\vec{\phi}=0$, where
$\vec{\phi}=(\phi_0,\phi_1,...,\phi_n)$:
\[\frac{\partial H_i}{\partial \phi_0}=2\pi \omega,\,i\in \{1,2,...,n\}.\]
\[\frac{\partial H_i}{\partial \phi_i}=\int_0^{2\pi}\Big(-\frac{g`(z_i)}{\omega}-2\frac{h`(0)}{\omega}\Big)d\theta,\,i\in \{1,2,...,n\}\]
Note that $g$ is a $2\pi$-periodic function, thus
\[\int_0^{2\pi}\Big(-\frac{g`(z_i(\theta))}{\omega}d\theta\Big)=\int_{z_i(0)}^{z_i(0)+2\pi}\Big(-\frac{g`(z_i(\theta))}{\omega}dz_i\Big)=-\frac{g(z_i(0)+2\pi)-g(z_i(0))}{\omega}=0.\]
Then
\[\frac{\partial H_i}{\partial \phi_i}=-2\frac{2\pi h`(0)}{\omega},\,i\in \{1,2,...,n\}.\]
Besides,
\[\frac{\partial H_{i}}{\partial \phi_m}=\frac{2\pi h`(0)}{\omega}(\delta(m-i-1)+\delta(m-i+1)),\]
where  $m,i\in\{1,2,...,n\}$, $m\neq i$  and
\[\delta(s)=
\begin{cases}
1,& \quad s=0\\
0, & \quad s\neq 0.
\end{cases}\]
Finally, we obtain the Jacobian matrix for $\vec{H}$
\begin{equation}
\frac{\partial \vec{H}}{\partial \vec{\phi}}=\frac{2\pi h`(0)}{\omega}
\begin{bmatrix}
a & -2 &1 & & & &  1\\
a & 1& -2 &1 & & & \\
a &  & 1  & -2&1& &\\
\vdots& & &\ddots& \ddots &\ddots&\\
a & & &&  1&-2 &1 \\
a & 1 & & & & 1 & -2
\end{bmatrix},
\end{equation}
where $a=\frac{\omega^2}{h`(0)}$. This is an $n\times (n+1)$ matrix
with $n$ linearly independent column vectors, thus
\[\text{rank}\Big(\frac{\partial \vec{H}}{\partial \vec{\phi}}\Big)=n\,,\]
which is just what the uniqueness condition requires for the
existence of periodic solutions. According to the theorem of
existence of periodic solutions, periodic solutions exist in our
general model and the Poincar\'{e}-Lindstedt method can be applied.

\section{Response to spatiotemporally periodic driving force}
\label{app:force}

On a circle network, with diffusive coupling and subject to a
spatiotemporal periodic driving, the equation of motion has the form
\begin{equation}\label{eq:general_driving}
\dot{f_i}=F(i,t)+d(f_{i-1}+f_{i+1}-2f_i)\,,
\end{equation}
where $F(i,t)=F(i+n,t)=F(i,t+T)$ with $n$ being the number of nodes
and $T$ the period of the driving. With some minor assumptions, a
periodic solution to this equation would be stable if $d>0$, because
the linearized equation takes the form of Eq.~(\ref{eq:stability2}).
In the following, we are mainly interested in the type of equations
that are relevant to the proof in section~\ref{sect:univ}.

Firstly, consider
\[F(i,t)=a\exp\left(j\Big(mt-\frac{(i-1)2km\pi}{n}\Big)\right)\,,\]
where $j$ is the imaginary number unit. Suppose the solution takes
the form
\begin{equation}
f_i(t)=A\exp\left(j\Big(mt-m(i-1)2k\pi/n\Big)\right)\,,
\label{eq:base}
\end{equation}
 and substituting it into Eq.~(\ref{eq:general_driving}) results in
\[A=\frac{a}{jm+4d\sin^2(\frac{mk\pi}{n})}\,.\]

With this basic solution, we can easily calculate solutions for
other types of driving terms. Consider
 \begin{equation}
 F(i,t)=a\sin\Big(mt-\frac{(i-1)m2k\pi}{n}\Big)\,.\label{eq:sin}
 \end{equation}
An implementation of Eq.~(\ref{eq:base}) leads to
\begin{equation}
\begin{split}
f_i(t)=&\frac{4ad\sin^2(mk\pi/n)}{m^2+(4d\sin^2(mk\pi/n))^2}\sin\Big(mt-\frac{m(i-1)2k\pi}{n}\Big)\\
      &-\frac{am}{m^2+(4d\sin^2(mk\pi/n))^2}\cos\Big(mt-\frac{m(i-1)2k\pi}{n}\Big) \,.
      \end{split}
      \label{eq:gen1}
\end{equation}
The cosine driving
\[F(i,t)=a\cos\Big(mt-\frac{(i-1)2mk\pi}{n}\Big)\,,\]
gives then
\begin{equation}
\begin{split}
f_i(t)=&\frac{am}{m^2+(4d\sin^2(mk\pi/n))^2}\sin\Big(mt-\frac{m(i-1)2k\pi}{n}\Big)\\
    &+\frac{4ad\sin^2(mk\pi/n)}{m^2+(4d\sin^2(mk\pi/n))^2}\cos\Big(mt-\frac{m(i-1)2k\pi}{n}\Big)\,.
      \end{split}
      \label{eq:gen2}
\end{equation}
All the solutions obtained above are  stable nonlinear
waves on the circle network, driven by spatiotemporally periodic
force.

\end{document}